\documentclass[twocolumn]{aastex631}

\usepackage{graphicx}
\usepackage{amssymb}
\usepackage{amsmath}
\usepackage{rotating}
\usepackage{pifont}
\newcommand{\cmark}{\ding{51}}%
\newcommand{\xmark}{\ding{55}}%

\def\lya{Ly$\alpha$}
\def\NHI{N$_{\rm H\,\textsc{I}}$}

\makeatletter
\DeclareRobustCommand{\HI}{%
  \mbox{H\,\check@mathfonts\fontsize\sf@size\z@\selectfont I}%
}
\makeatother

\makeatletter
\DeclareRobustCommand{\MgII}{%
  \mbox{Mg\,\check@mathfonts\fontsize\sf@size\z@\selectfont II}%
}
\makeatother

\makeatletter
\DeclareRobustCommand{\HII}{%
  \mbox{H\,\check@mathfonts\fontsize\sf@size\z@\selectfont II}%
}
\makeatother

\received{November 2, 2023}
\revised{November 27, 2023}
\accepted{November 29, 2023}

\shorttitle{A survey of Ly$\alpha$ emission around Damped Ly$\alpha$ absorbers at $z \approx 2$ with KCWI}
\shortauthors{Oyarz\'un et al.}

\graphicspath{{./}{figures/}}

\begin{document}

\title{A survey of Ly$\alpha$ emission around Damped Ly$\alpha$ absorbers at $z \approx 2$\\ with the Keck Cosmic Web Imager}

\correspondingauthor{Grecco A. Oyarz\'un}
\email{goyarzu1@jhu.edu}

\author{Grecco A. Oyarz\'un}
\affiliation{The William H. Miller III Department of Physics \& Astronomy, Johns Hopkins University, Baltimore, MD 21218, USA}

\author{Marc Rafelski}
\affiliation{Space Telescope Science Institute, 3700 San Martin Drive, Baltimore, MD 21218, USA}
\affiliation{The William H. Miller III Department of Physics \& Astronomy, Johns Hopkins University, Baltimore, MD 21218, USA}

\author{Nissim Kanekar} 
\affiliation{National Centre for Radio Astrophysics, Tata Institute of Fundamental Research, Pune University, Pune 411007, India}

\author{J. Xavier Prochaska}
\affiliation{Department of Astronomy \& Astrophysics, UCO/Lick Observatory, University of California, 1156 High Street, Santa Cruz, CA 95064, USA}
\affiliation{Kavli Institute for the Physics and Mathematics of the Universe (Kavli IPMU), 5-1-5 Kashiwanoha, Kashiwa, 277-8583, Japan}

\author{Marcel Neeleman}
\affiliation{National Radio Astronomy Observatory, 520 Edgemont Road, Charlottesville, VA, 22903, USA}

\author{Regina A. Jorgenson} 
\affiliation{Maria Mitchell Observatory, Nantucket, MA, United States}



\begin{abstract}
We present Keck Cosmic Web Imager Ly$\alpha$ integral field spectroscopy of the fields surrounding 14 Damped \lya\ absorbers (DLAs) at $z \approx 2$. Of these 14 DLAs, nine have high metallicities ([M/H]~$> -0.3$), and four of those nine feature a CO-emitting galaxy at an impact parameter~$\lesssim 30$~kpc. Our search reaches median \lya\ line flux sensitivities of $\sim$$2 \times 10^{-17}$~erg~s$^{-1}$~cm$^{-2}$ over apertures of $\sim$6~kpc and out to impact parameters of $\sim$50~kpc. We recover the \lya\ flux of three known  Ly$\alpha$-emitting \HI-selected galaxies in our sample. In addition, we find two \lya\ emitters at impact parameters of $\approx 50-70$~kpc from the high metallicity DLA at $z \approx 1.96$ toward QSO B0551-366. This field also contains a massive CO-emitting galaxy at an impact parameter of $\approx 15$~kpc. Apart from the field with QSO B0551-366, we do not detect significant \lya\ emission in any of the remaining eight high-metallicity DLA fields. Considering the depth of our observations and our ability to recover previously known Ly$\alpha$ emitters, we conclude that \HI-selected galaxies associated with high-metallicity DLAs at $z \approx 2$ are dusty and therefore might feature low \lya\ escape fractions. Our results indicate that complementary approaches --- using Ly$\alpha$, CO, H$\alpha$, and [C\,{\sc ii}]~158$\mu$m emission --- are necessary to identify the wide range of galaxy types associated with $z \approx 2$ DLAs.
\end{abstract}


\section{Introduction}
\label{1}

The \HI\ cycle within and around galaxies is a critical component in our models of galaxy formation and evolution. We know that galaxies must acquire \HI\ from the intergalactic medium in order to sustain their star-formation (\citealt{prochaska2005,keres2005,walter2020}). The inflow of \HI\ is counteracted by outflows of metal-enriched gas powered by active galactic nuclei and/or the late stages of stellar evolution, thereby regulating the rate at which galaxies form their stars (e.g. \citealt{keres2005,tumlinson2017}). Furthermore, the removal of \HI\ from galaxies through environment-driven processes (\citealt{gunn-gott1972,kawata-mulchaey2008,cortese2021}) is often invoked to explain the quenched fractions and assembly histories of satellite galaxies \cite[e.g.][]{pasquali2010,wetzel2013,gallazzi2020,trussler2021,werle2022,oyarzun2023}.

To characterize the \HI\ content in and around galaxies at low redshift, we often turn to studies of 21\,cm emission \citep[e.g.][]{verheijen2001,walter2008,begum2008,chung2009,heald2011,catinella2018}. Single-dish 21\,cm studies have yielded \HI\ emission-line detections for thousands of galaxies at low redshifts (e.g. \citealt{zwaan2005,haynes2018}), while 21\,cm mapping studies have been used to, for example, quantify the star-formation efficiency in nearby galaxies (\citealt{leroy2008}), measure the sizes of \HI\ disks \citep[e.g.][]{wang2016b}, determine galaxy rotation curves \citep[e.g.][]{walter2008,begum2008}, and search for extra-planar gas \citep[e.g.][]{heald2011}. However, the faintness of the 21\,cm transition has so far prevented us from detecting \HI\ in emission from individual galaxies beyond $z \approx 0.4$ \citep{fernandez2016} or in stacked imaging beyond $z \approx 1.4$ \citep{chowdhury2020,chowdhury2021,chowdhury2022}.

Alternatively, absorption signatures in the spectra of background quasars (QSOs) produced by high \HI\ column density gas \citep[Damped Ly$\alpha$ absorbers, or DLAs;][]{wolfe2005} remain the quintessential technique for studying \HI\ at high redshift. Through DLA characterization, we have been able to constrain the column densities, metallicities, kinematics, dust depletion, molecular fractions, and gas temperatures of \HI\ reservoirs up to $z\sim 5.5$ (e.g. \citealt{balashev2017,prochaska-wolfe1997,kanekar2014,noterdaeme2008,neeleman2013,neeleman2015,klimenko2020}). Moreover, DLAs have been instrumental in determining the evolution of the metal enrichment and cosmic \HI\ mass density since $z\approx 5$ \citep[e.g.][]{noterdaeme2012,rafelski2012,rafelski2014,jorgenson2013,prochaska2013,crighton2015,rao2017}, connecting with estimates at lower redshifts from 21\,cm observations \citep[e.g.][]{jones2018,bera2019}.

On the other hand, it has been challenging to associate the properties of \HI\ measured in absorption with the properties of galaxies measured in emission. DLA galaxies are typically much fainter than the background QSOs, and are found over a wide range of impact parameter ($b$), which makes the identification of the galaxy through standard optical imaging and spectroscopy very challenging. Despite many searches, only about a dozen galaxies associated with DLAs at $z \gtrsim 2$ were detected in over a quarter of a century \citep[e.g.][]{moller-warren1993,fumagalli2015}. Fortunately, the detection rate has since increased. After it was realized that the stellar mass and gas-phase metallicity relation (e.g. \citealt{tremonti2004}) also holds for absorption-selected systems \citep[e.g.][]{moller2004,ledoux2006,fynbo2008}, studies in the rest-frame UV/optical have started to target high-metallicity DLAs ([M/H]$\gtrsim -1.3$) with great success \citep[e.g.][]{krogager2017}. 

At the same time, the advent of the Atacama Large Millimeter/sub-millimeter Array (ALMA) has enabled a search for DLA galaxies at millimeter and sub-millimeter wavelengths, where the QSOs are much fainter and line emission from cool or cold gas can be luminous. ALMA and Northern Extended Millimetre Array (NOEMA) images have been used to identify a further dozen star-forming counterparts of high-metallicity ([M/H]$>-1.3$) DLAs through their CO emission at $z\approx 2$ or their [C\,{\sc ii}]~158$\mu$m emission at $z\approx4$ \citep{neeleman2017,neeleman2018,fynbo2018,neeleman2019,kanekar2020,kaur2022a}.

Although searches for DLA galaxies in the CO and [C\,{\sc ii}]~158$\mu$m transitions have shown to be quite efficient, the downside is that they are bound to miss galaxies with high CO-to-H$_2$ conversion factors or with relatively low molecular gas masses \citep[$\lesssim 1-5 \times 10^{10}$ M$_{\odot}$;][]{neeleman2019,kanekar2020,kaur2022a}. Such undetected galaxies could reside at lower impact parameters than mm-detected galaxies, perhaps explaining why the impact parameters between mm-detected galaxies and DLA sightlines --- $b < 30$~kpc at $z\approx 2$ and $b\gtrsim 25$~kpc at $z\approx 4$ \citep{neeleman2017,neeleman2018,neeleman2019,kanekar2020} --- can exceed the sizes of \HI\ gas reservoirs at $z\approx 2-4$ in simulations ($<30$~kpc; \citealt{rhodin2019,stern2021}).

To search for galaxies with high CO-to-H$_2$ conversion factors and/or low molecular gas masses, we can turn to rest-frame UV/optical emission. Among the standout emission lines at these wavelengths is \lya, which is produced in \HII\ regions by recombining \HI\ gas. However, only $\approx 20-25\%$ of Lyman-break galaxies (LBGs; \citealt{steidel1996,steidel1999,steidel2003,shapley2003,stark2009}) at $z\approx 2$ show \lya\ emission \citep[e.g.][]{cassata2015}. This is presumably due to the ease with which \lya\ is scattered by \HI\ and absorbed by dust grains \citep[e.g.][]{dijkstra2012,duval2014,rivera-thorsen2015,gronke2016,gronke2016b}. As a result, the equivalent width of the \lya\ line is particularly high in galaxies with low \HI\ gas covering fractions and low dust extinctions, i.e., low stellar-mass (M$_* \lesssim 10^{10}$ M$_{\odot}$) galaxies \citep[e.g.][]{oyarzun2016,oyarzun2017}.

For these reasons, \lya\ searches have been exploited to detect DLA galaxies down to much lower metallicities ([M/H]~$\lesssim-1.3$) than mm-wave transitions \citep[e.g.][]{moller2004,mackenzie2019,lofthouse2023}. To search for \lya\ emission at the DLA redshift, early searches used narrow-band imaging, yielding a few tentative detections \citep[e.g.][]{smith1989,lowenthal1991,giavalisco1994,fynbo2003,kulkarni2006,grove2009,fynbo2023}. More recently, \lya\ searches have capitalized on the sensitivities of new state-of-the-art optical integral field units \citep[IFUs; e.g.][]{christensen2007,peroux2011,peroux2012,fumagalli2017,mackenzie2019,lofthouse2023}. For instance, \citet{fumagalli2017} used the Multi Unit Spectroscopic Explorer (MUSE) on the Very Large Telescope (VLT) to identify three Ly$\alpha$ emitting sources associated with a DLA at $z\sim 3.25$. Building on this result, subsequent surveys with VLT/MUSE have found more than 20 \lya\ emitters out to impact parameters of $\approx 300$~kpc in the fields of $\approx 15$ low-metallicity DLAs at $z \approx 3$ \citep[e.g.][]{mackenzie2019,lofthouse2020,lofthouse2023}.

Motivated by the success of recent searches in \lya\ with IFUs, in this work we searched for \lya\ emission in the fields of 14 DLAs --- nine of which have been previously studied in mm-wave CO emission --- with the Keck Cosmic Web Imager Integral Field Spectrograph \citep[KCWI;][]{morrissey2018} on the Keck~II telescope. KCWI is an outstanding instrument for this search because of its high sensitivity at the observed-frame wavelength of $z\approx 2$ \lya\ emission ($\lambda \approx 4500$\,\AA). Moreover, the effective field-of-view of $\sim 100$~kpc enables us to cover the impact parameter range expected for the primary emission counterparts of high \HI\ column density  absorbers \citep[$\lesssim 30$~kpc; e.g.][]{rahmati-schaye2014,rhodin2019}.

The paper is structured as follows. We define our target sample of DLAs and describe the observations in Sections \ref{2} and \ref{3}. We detail our methodology to identify and characterize Ly$\alpha$ emission in Section~\ref{4}. We present our results in Section~\ref{5}, discuss their interpretation in Section~\ref{6}, and provide a summary of the paper in Section~\ref{7}. Throughout the paper, we assume $H_0=70$~km~s$^{-1}$~Mpc$^{-1}$. All magnitudes are reported in the AB system (\citealt{oke1983}). 

\begin{figure*}[t]
\centering
\includegraphics[width=7.1in]{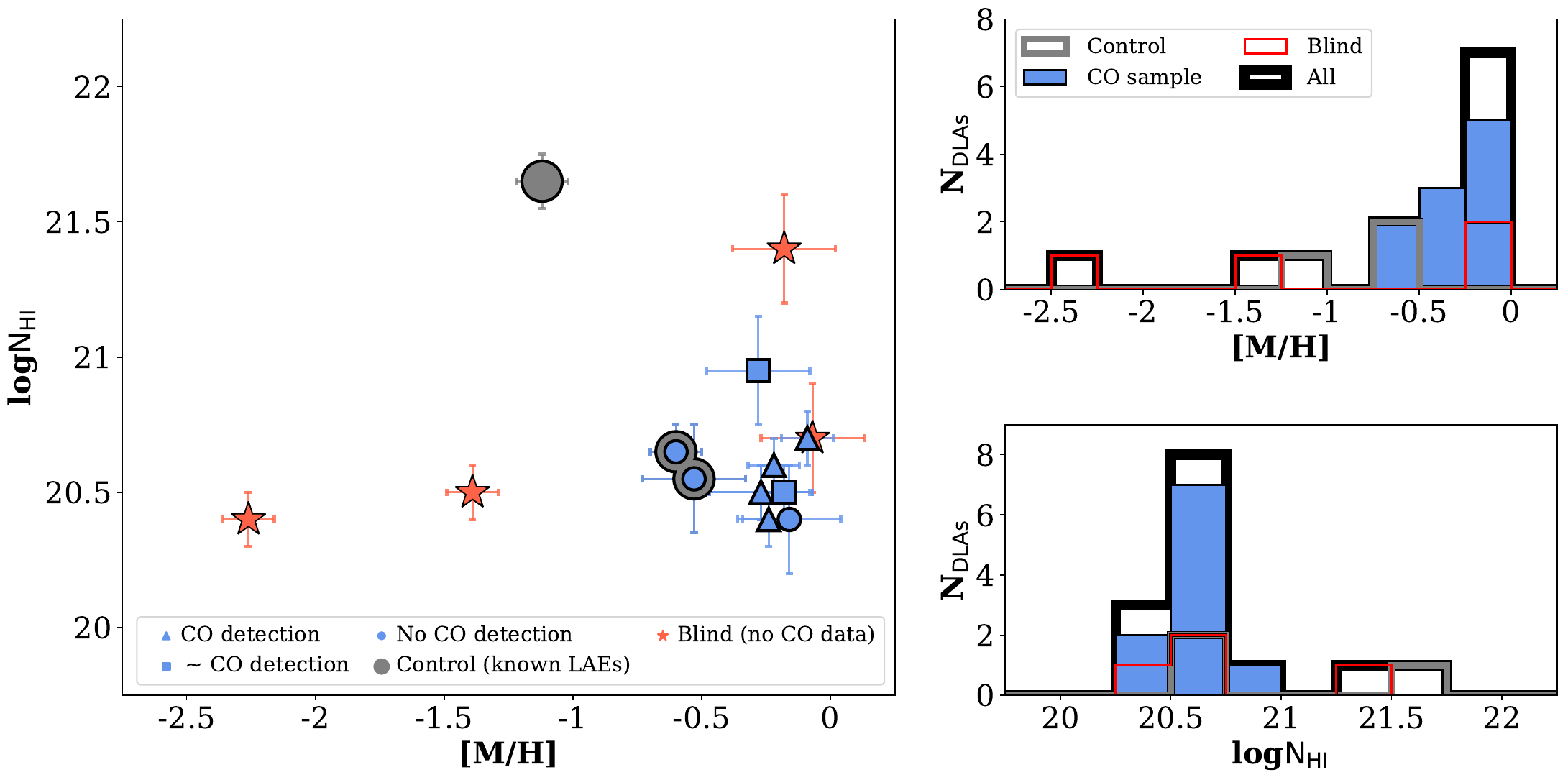}
\caption{The sample. \textbf{Left:} Distribution of the DLAs in metallicity-\NHI\ space. Control DLA galaxies (i.e., known \lya\ emitters) are plotted as large gray circles. DLAs previously targeted for CO observations are denoted with the small blue markers. Nondetections in CO are plotted as circles, potential CO detections as squares, and CO detections as triangles. DLAs without mm-observations are plotted as red stars. \textbf{Right:} Histograms of the metallicities and \HI\ column densities of our 14 DLAs. The full sample is plotted in black, control DLA galaxies in gray, DLAs with CO observations in blue, and DLAs without known prior \lya\ or CO observations in red.}
\label{fig:sample}
\end{figure*}

\section{The Sample}
\label{2}

\citet{rafelski2012,rafelski2014} employed the Echellette Spectrograph and Imager (ESI; \citealt{Keck-ESI}) and the High Resolution Echelle Spectrometer (HIRES; \citealt{Keck-HIRES}) on the Keck Telescopes to obtain high-resolution spectroscopy for 50 QSOs. Novel in their analysis was that the DLA selection was based solely on \HI\ column density (\NHI), thus avoiding any metallicity biases. As part of their work, they also reanalyzed a collection of QSO spectra from the literature selected not to have a metallicity bias. It is from this sample that the majority of the DLAs in our work originate (B0458-020, J2206-1958a, B1228-113, B0201+365, B0551-366, B1230-101, J0453-1305, and J2206-1958b). In addition, three high-metallicity DLAs in our sample --- J1305+0924, Q1755+578, and J1013+5615 --- were characterized by \citet{berg2015}. Completing our sample are J2222-0946, J2225+0527, and J1709+3258, which were characterized by \citet{fynbo2010}, \citet{krogager2016}, and \citet{kaplan2010}, respectively. The properties of our DLAs --- i.e., redshifts, \NHI, and metallicities--- are presented in Table~\ref{table:data} and in Figure~\ref{fig:sample}. As is clear from the figures, our 14 target DLAs are dominated by high-metallicity absorbers ([M/H]~$>-0.3$).


\begin{deluxetable*}{ccccccc}
\tabletypesize{\normalsize}
\tablecaption{Sample. The columns correspond to the QSO sightline identifier (1), QSO right ascension (2), QSO declination (3), magnitude of the QSO in the g-band measured by the \citet{gaiadr3} (4), redshift of the QSO (5), DLA subsample (6), and exposure time (7).}
\setlength{\tabcolsep}{0.19in}
\tablehead{
\colhead{QSO} & \colhead{ra} & \colhead{dec} & \colhead{$m_{\rm qso}$} & \colhead{$z_{\rm qso}$} & \colhead{Subsample} & \colhead{Exp. time [hr]}
}
\startdata 
B0458-020 & 05:01:12.80	& -01:59:14.25 & 18.69 & 2.29 & control & 1\\
J2206-1958a$^{\dagger}$ & 22:08:52.07 & -19:43:59.86	& 16.82 & 2.57 & control/CO & 1.5\\
J2222-0946 & 22:22:56.11 & -09:46:36.28 & 18.04 & 2.93 & control/CO & 1\\
J1305+0924 & 13:05:42.77 & 09:24:27.75 & 18.95 & 2.04 & CO& 1\\
B1228-113 & 12:30:55.56 & -11:39:09.79 & 19.63 & 3.53 & CO & 1\\
B0201+365 & 02:04:55.60	& 36:49:17.99 & 18.04 & 2.91 & CO & 1\\
B0551-366 & 05:52:46.18 & -36:37:27.60 & 16.98 & 2.32 & CO & 0.83\\
J2225+0527 & 22:25:14.70 & 05:27:09.06 & 17.84 &  2.32 & CO & 1\\
J1709+3258 & 17:09:09.28 & 32:58:03.40 & 19.17 & 1.89 & CO & 0.5\\
B1230-101 & 12:33:13.16 & -10:25:18.44 & 19.26 & 2.39 & CO & 0.83\\
J0453-1305 & 04:53:13.57 & -13:05:55.09 & 16.99 & 2.26 & blind & 0.66\\
J2206-1958b$^{\dagger}$ & 22:08:52.07 & -19:43:59.86 & 16.82 & 2.57 & blind & 1.5\\
Q1755+578 & 17:56:03.63 & 57:48:47.99 & 18.55 & 2.11 & blind & 1.17\\
J1013+5615 & 10:13:36.38 & 56:15:36.41 & 18.61 & 3.63 & blind & 1\\
\enddata
\tablecomments{There are two DLAs along the sightline of J2206-1958 ($\dagger$).}
\label{table:data}
\vspace{0em}
\end{deluxetable*}

\subsection{The CO subsample}
By design, our sample of target DLAs is dominated by systems with ancillary CO observations. These observational campaigns were conducted with ALMA  \citep{kanekar2020}, NOEMA \citep{kaur2022a}, and the Karl G. Jansky Very Large Array (JVLA; B. Kaur et al. 2023, in preparation). Those authors searched for mm-wave redshifted CO(1--0), CO(2--1), CO(3--2), or CO(4--3) emission from the fields of 20 high-metallicity ([M/H]$\gtrsim -1.0$) DLAs at $z \approx 1.7-2.6$. They identified six robust ($> 5\sigma$ significance) and two tentative ($4-5\sigma$) detections of redshifted CO emission, obtaining a $\approx 50\%$ CO detection rate of galaxies around their highest-metallicity DLAs ([M/H]~$\gtrsim -0.35$). Five of the CO detections have impact parameters in the range $b = 5-30$~kpc \citep{neeleman2018,kanekar2020,kaur2022a}. The sixth CO detection is at $b \approx 100$~kpc \citep{fynbo2018,kanekar2020} in a system at $z\approx 2.58$ where a DLA galaxy has been identified at $b \approx 16$~kpc (Q0918+1636; \citealt{fynbo2011,fynbo2013}). For the 13 CO nondetections, the median 3$\sigma$ upper limit on the  molecular gas mass is $\approx 1.2 \times 10^{10}$ M$_{\odot}$, for a CO-to-H$_2$ conversion factor of $\alpha_{\rm CO} = 4.36$~M$_\odot$~(K~km~s$^{-1}$~pc$^2$)$^{-1}$ \citep[e.g.][]{bolatto2013}. Of the 14 DLAs in our sample, nine have CO studies, with four CO detections, two tentative detections, and three nondetections see (Tables \ref{table:data} and \ref{table:results}). We will refer to these nine DLAs as the CO subsample throughout.

\subsection{The Control subsample}

Earlier \lya\ searches for galaxies at low impact parameters have obtained detections in two of the DLAs in the CO subsample. \lya\ emission in the field of the $z = 2.3543$ DLA towards J2222-0946 was first reported, at $b\approx 6$~kpc, by \citet{fynbo2010} \citep[see also][]{,krogager2013,krogager2017}. The $z = 1.9200$ DLA towards J2206-1958 also has an associated \lya-emitting galaxy (at $b \lesssim 6$ kpc; \citealp{moller2002}). No CO emission was detected in either of these fields by \citet{kanekar2020}. A third DLA in our sample, this time without ancillary CO observations, has also been detected in \lya. This galaxy is at $z=2.0395$ towards B0458-020 and at an impact parameter of $b\lesssim 3$~kpc \citep{moller2004,krogager2017}. Together, the three known \lya\ emitters in the fields of J2222-0946, J2206-1958, and B0458-020 constitute our control sample. They were used to assess the efficacy of our survey and to search for additional \lya\ associations at larger impact parameters. 

\subsection{The ``Blind'' subsample}

The remaining four DLA fields (Table \ref{table:data})
compose the ``blind'' sample, i.e. absorbers without earlier searches for the associated galaxies. One of these four DLAs lies towards J2206-1958, a sightline containing one of our control DLAs. The remaining three blind-sample DLAs were observed due to their convenient celestial coordinates in the context of our observational strategy. They will be the targets of future CO observations with ALMA , NOEMA, and/or the JVLA.

\begin{figure*}
\centering
\includegraphics[width=7.1in]{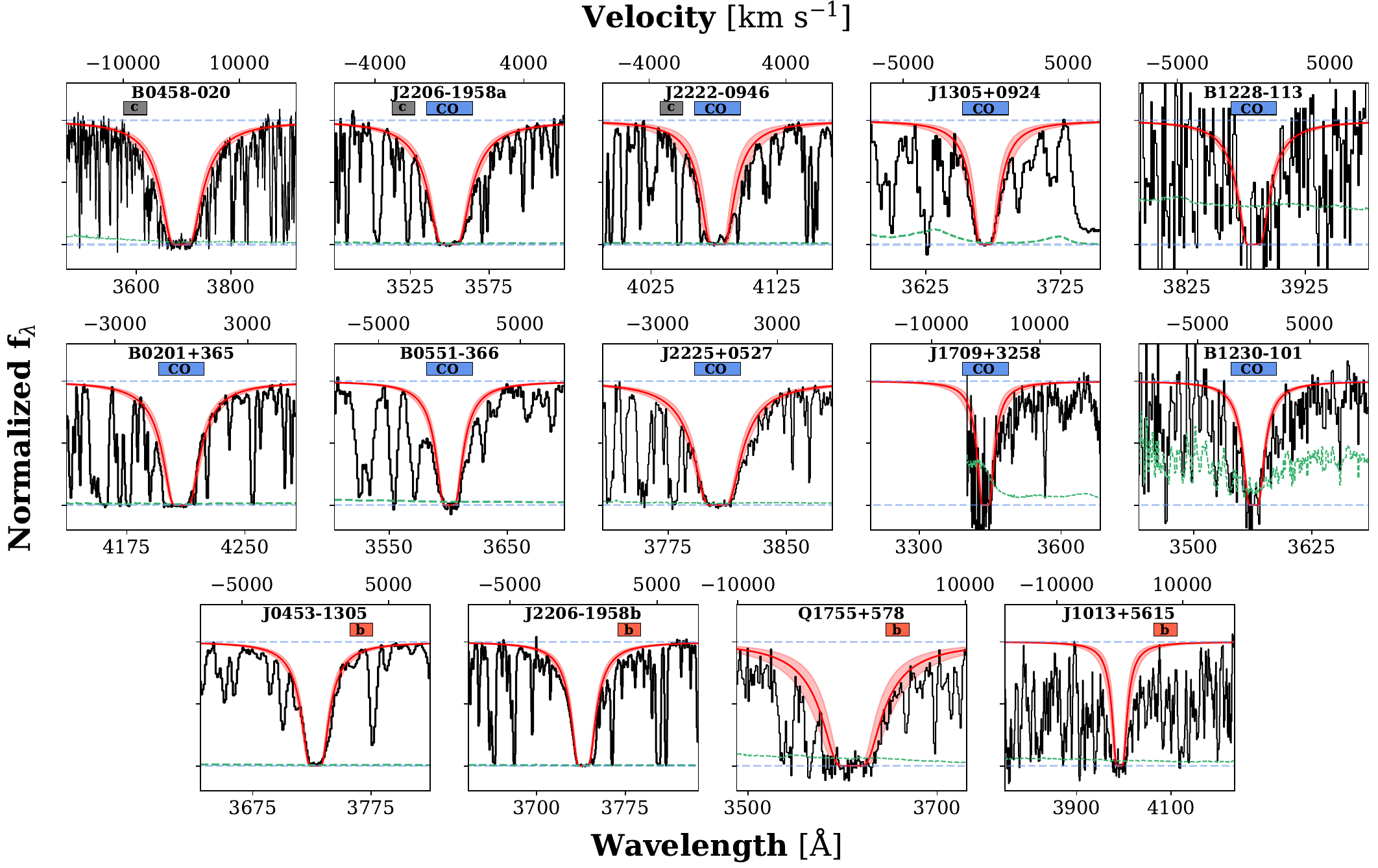}
\caption{Spatially integrated KCWI spectra for the 14 QSOs after continuum normalization. Every panel shows the spectrum (continuous black lines) and the error (dashed green lines) at the wavelength of the DLA. The red lines and shaded regions show the best solution and the error on the DLA fits that were used to determine the redshifts and \HI\ column densities of the absorbers (\citealt{rafelski2012,rafelski2014}). The symbols underneath the name of the target denote the sample the DLAs belong to. Note that the DLA along the J1305+0924 sightline is proximate, i.e., the DLA is within 5000~km~s$^{-1}$ of the \lya\ emission from the QSO.}
\label{fig:dla_spectra}
\end{figure*}

\newpage

\section{Observations}
\label{3}

The 14 DLA fields in this work were observed with the Keck Cosmic Web Imager Integral Field Spectrograph \citep[KCWI;][]{morrissey2018}, on the Keck~II  telescope. KCWI is optimized for observations in the $3500-5600$\,\AA\ spectral range with resolution $R=1000-20,000$. The size of a spaxel is $\sim 0.\arcsec 7$, which corresponds to a physical size of $\sim 6$~kpc at $z\sim 2$. We opted for the configuration with a $20 \arcsec \times 16 \arcsec$ field of view at a spatial resolution of $0.\arcsec 7$ and with a spectral resolution of $R\sim 2000$ or $R\sim 4000$, depending on the target. At $z\sim2$, this configuration corresponds to a field-of-view of $\approx 170~{\rm kpc} \, \times 130$~kpc and a spatial resolution of $\approx 6$~kpc. 

The first set of KCWI observations was conducted in 2019 September and October, with later observing runs in 2021 February and April. The on-target exposure time varied between 0.5~hr and 1.5~hr, with DLAs with ancillary CO observations given priority. We obtained line flux sensitivities of $\approx 0.1 - 5 \times 10^{-16}$~erg~s$^{-1}$~cm$^{-2}$ at $S/N=5$, depending on the exposure time and the DLA redshift (i.e. the redshifted \lya\ wavelength). 

The KCWI data were analysed with PypeIt\footnote{\href{https://pypeit.readthedocs.io/en/release/}{https://pypeit.readthedocs.io/en/release/}}, a Python package designed for the semi-automated reduction of astronomical spectroscopic data\footnote{\href{https://pypeit.readthedocs.io/en/release/spectrographs/keck_kcwi.html}{https://pypeit.readthedocs.io/en/release/spectrographs/\\keck\_kcwi.html}}. We used the built-in routines to perform bias subtraction, dark correction, and trace pattern identification. For wavelength calibration, we used FeAr lamps. The flat-fielding of the data accounted for pixel-by-pixel variations and the dome illumination pattern. The faintness of our targets allowed PypeIt to use the science frames themselves to perform sky subtraction. Bright objects, such as the QSO in each field, were detected and masked in the computation of the sky model. The astrometric solution of the data cubes was revised using the coordinates of the QSOs.

\subsection{Flux calibration}
\label{2.3}
Observations of standard stars ($m\sim10$) prior to or after the science exposures were used for flux calibration. We used PypeIt to compute the flux sensitivity curves from these standard stars. The first step of the process was to flux calibrate the data cubes of the standard stars. We verified that this step was performed appropriately by comparing the flux-calibrated standard data cubes with the well-characterized standard spectra. Then, the flux-calibrated standard star data cubes were used as the input sensitivity curves in the co-addition of all the data cubes (i.e., exposures) for each target. 

To quantify the accuracy of our flux calibration, we searched for publicly available observations of the QSOs ($m\sim 18$; Table \ref{table:data}) in our sample. Four of the QSOs --- J2222-0946, J1305+0924, J1709+3258, and J1013+5615 --- were observed in the Baryon Oscillation Spectroscopic Survey (BOSS) of the Sloan Digital Sky Survey (SDSS; \citealt{dr12}). We found that our flux-calibration is consistent with that of SDSS/BOSS within a factor of $\sim 2$. To maximize the accuracy of our flux calibration, we estimated a flux-correction factor by maximizing the likelihood between the KCWI and the SDSS/BOSS spectra. An optimal flux correction factor of 1.8 was found and used to re-scale our flux-calibrated spectra. The dispersion among these measurements indicates that our flux calibration has a standard error of 12\%. All errors on the line fluxes reported in this paper include this factor of 0.12. Finally, all fluxes were corrected for Milky Way extinction using the \citet{fitzpatrick1999} extinction curve, the dust reddening maps from \citet{schlegel1998}, and the reddening to $A_V$ conversion tabulated in \citet{schlafly-finkbeiner2011}. 

\newpage

\section{Methodology}
\label{4}

\subsection{Subtraction of the sky background and the QSO continuum}
\label{3.1}

Figure \ref{fig:dla_spectra} shows the fully-reduced KCWI spectra towards each target QSO centered at, and zoomed in on, the redshifted \lya\ absorption line of each of the 14 DLAs. Before searching for emission lines at the DLA redshift, characterization of the QSO emission was required. We started by constructing a QSO continuum model for the data cube of every target. To do this, we computed $\phi_{qso}(\lambda)$, a least-squares Chebyshev series fit to the QSO continuum in the brightest spaxel. Then, $\phi_{qso}(\lambda)$ was scaled throughout the data cube to produce the initial continuum model cube
\begin{equation}
\mathcal{C}_{qso}(x,y,\lambda) = A(x,y)\phi_{qso}(\lambda),
\end{equation}
where $A(x,y)$ is the spaxel-dependent QSO continuum scaling factor. The spaxel-dependent error in the spectra --- $\sigma(x,y,\lambda)$ --- was then used to estimate
\begin{equation}
\mathcal{SN}_{qso}(x,y) = \frac{1}{n} \sum^{n}_{\lambda} \frac{\mathcal{C}_{qso}(x,y,\lambda)}{\sigma(x,y,\lambda)},
\end{equation}
i.e., the mean signal-to-noise of the QSO continuum throughout the data cube. Thresholds in $\mathcal{SN}_{qso}(x,y)$ were imposed to identify spaxels affected by the QSO continuum. Depending on the target, we found the optimal value for the threshold --- determined through visual inspection --- to vary between 5 and 14.

In order to estimate the background emission throughout each cube, all spaxels affected by QSO emission were masked. After masking, the median spectrum across each data cube was fitted with a Chebyshev series to obtain the background model $\mathcal{C}_{sky}(\lambda)$. Implementation of the median instead of the mean ensured that no serendipitous background sources affected our estimate of $\mathcal{C}_{sky}(\lambda)$. For each field, a background-subtracted data cube was then computed, which then was used to measure $\mathcal{C}_{qso}(x,y,\lambda)$ again; this time without contributions from the sky background. Finally, the QSO continuum model was subtracted from this residual data cube.

\begin{figure}[t]
\centering
\includegraphics[width=3.4in]{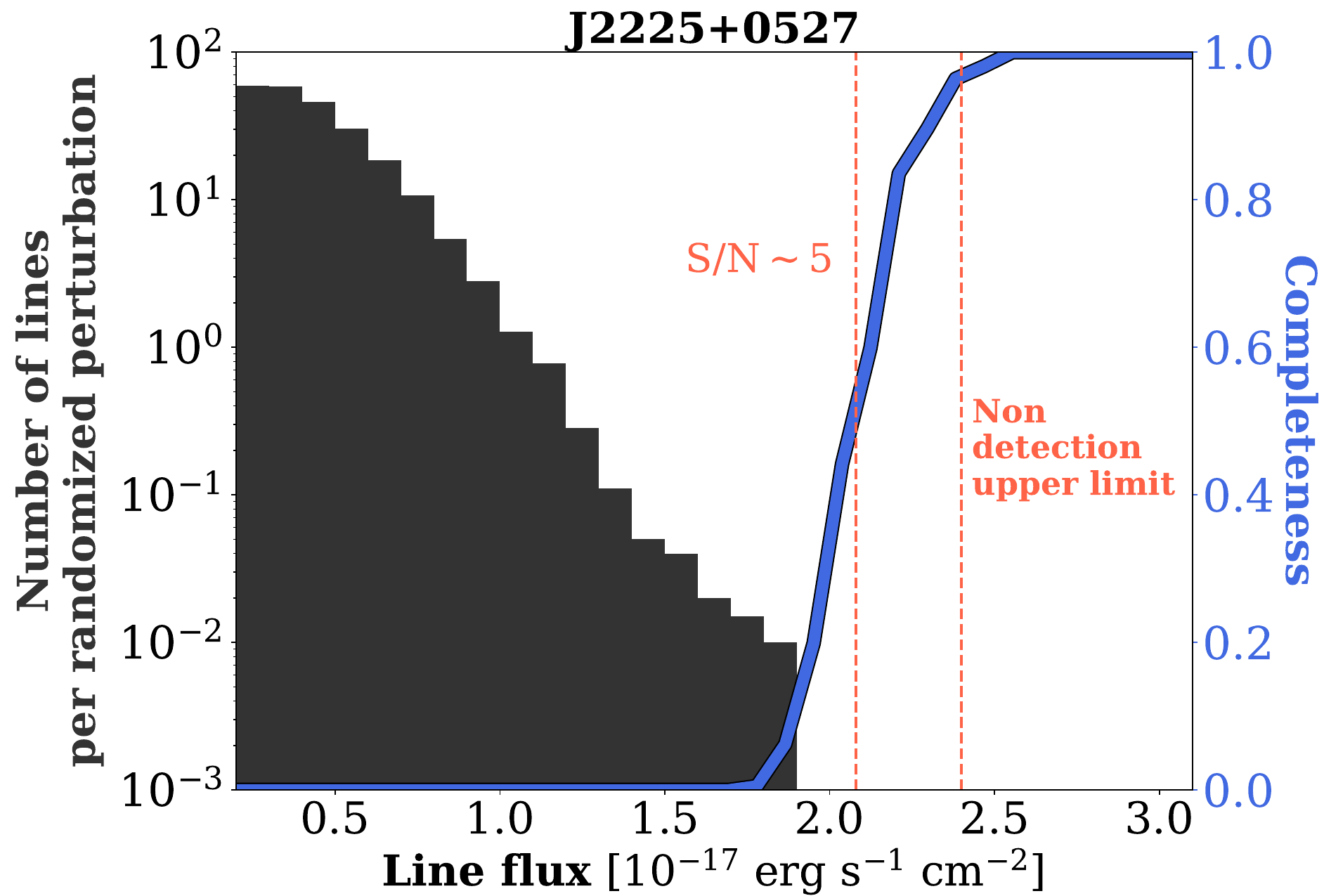}
\caption{Example of the output from our line detection routine for the DLA along QSO J2225+0527. Plotted are all false positives (black histogram, left y-axis) and our completeness (blue filled line, right y-axis). These quantities are shown at the wavelength of the DLA absorption and after collapsing over the spatial component. The false positive histogram was obtained by running our line detection code on error perturbed spectra. The completeness curve corresponds to the fraction of planted lines that we were able to measure with at least $S/N_{Ly\alpha} = 5$ as a function of line flux. The detection threshold and nondetection upper limit (completeness $\sim$ 95\%) are shown as red dashed lines.}
\label{fig:line_stats}
\end{figure}

\subsection{Ly$\alpha$ line measurements}
\label{3.2}

We searched for emission lines in the processed data cubes in 1000~km~s$^{-1}$ velocity slices centered at the wavelength of the DLA. To avoid the wings of the DLA, narrower velocity slices had to be used at the position of the QSO. Depending on the target, values between 500~km~s$^{-1}$ and $1000$~km~s$^{-1}$ were used. Then, gaussian emission lines were fitted in these velocity slices for every spaxel. The free parameters of the fit were the line centroid, the peak flux density, and the line width. This step was repeated 500 times on 500 error-perturbed spectra for every spaxel, which yielded a line flux distribution for every spaxel around the  expected redshifted \lya\ wavelength. The mean and the standard deviation of each distribution were taken as the measured line flux and line flux error (F$_{Ly\alpha}$ and $eF_{Ly\alpha}$) at every spaxel.

To associate a $S/N$ to an emission line, we quantified the line flux distribution across the data cube around the expected redshifted \lya\ wavelength. To do this, we first masked all emission lines with $\geq 2\sigma$ significance, and then added back the background and the QSO emission. We then produced 200 error-perturbed data cubes for every target and carried out background- and QSO-subtraction for each such cube. Line fluxes were then measured across these data cubes, yielding a false positive line flux distribution for every spaxel around the redshifted \lya\ wavelength (see Figure~\ref{fig:line_stats}). The line fluxes of the error-perturbed data cubes were then rank-ordered to obtain the correspondence between line flux and percentile. This correspondence was used to assign a $S/N$ to every Ly$\alpha$ line flux measurement in the original data cube.

We found that $S/N_{Ly\alpha} = 5$ effectively separates significant and spurious detections in the KCWI data cubes. For all significant detections, we recomputed the line fluxes to ensure that we account for spatially extended sources. To this end, we coadded the spectra of all the spaxels adjacent to the spaxel containing the detection. Our line flux fitting algorithm (see above) was then used to estimate the total line flux in the coadded spectrum. The Ly$\alpha$ line fluxes and errors reported throughout the rest of the paper correspond to the values measured in the coadded spectra. The $S/N$ was not recalculated, i.e., the values of $S/N_{Ly\alpha}$ reported throughout were measured in the spaxel showing the highest significance.

For targets with no significant detections, we characterized our line flux sensitivity within the KCWI field-of-view. To this end, we inserted 10,000 emission lines with varying line fluxes and profile shapes in randomized locations within each data cube. The shapes of the simulated emission lines were chosen to be Gaussian, with 1$\sigma$ widths between 150~km~s$^{-1}$ and 225~km~s$^{-1}$. The line flux at which we were able to recover 95\% of the lines with at least $S/N_{Ly\alpha} = 5$ was defined as our \lya\ sensitivity limit (Figure \ref{fig:line_stats}).

Inspection of the data cubes revealed 14 sources of continuum emission. With no line emission at the expected redshifted \lya\ wavelength, these sources are most likely foreground interlopers at redshifts lower than that of the DLA. Of the 14 sources, 7 are bright at the redshifted \lya\ wavelength, and we  therefore highlight them in white boxes in figures throughout the paper.

\begin{figure}
\centering
\includegraphics[width=3.4in]{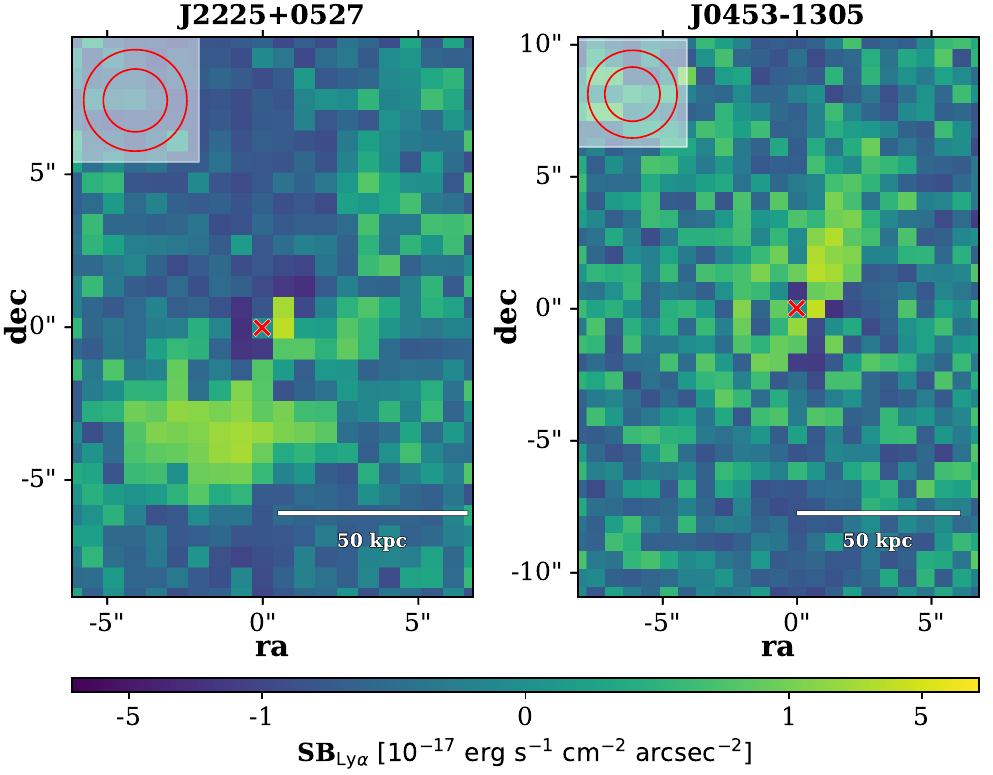}
\caption{Two spatially extended \lya\ nebulae around QSO~J2225+0527 and QSO~J0453-1305. Plotted is the surface brightness in \lya\ after subtraction of the QSO continuum. The sizes of the point spread functions ($1\sigma$ and $2\sigma$ contours) are shown in the top left corners. The spatial extension of the emission extends over at least two spatial beams, with the structure in QSO~J2225+0527 extending over at least 70~kpc and the nebulae around QSO~J0453-1305 apparent out to at least $40$~kpc.
}
\label{fig:nebulae}
\end{figure}

We also performed a search for \lya\ nebulae around the QSOs of our sample. This search could not be performed for four of the 13 QSOs (J2222-0946, B1228-113, B0201+365, and J1013+5615) because their \lya\ emission was redshifted out of the spectral coverage of KCWI. Out of the remaining nine QSOs, we found evidence for spatially extended \lya\ emission in two cases. The emission extends over at least 70~kpc around QSO~J2225+0527 and over $\approx 40$~kpc around QSO~J0453-1305. The two \lya\ nebulae are shown in Figure \ref{fig:nebulae}. In passing, we note that the spatial resolution of our observations at the redshift of the QSO is typically $\approx 14$~kpc FWHM, implying that the \lya\ emission extends over at least two spatial beams in both cases. Similarly to \citet{herenz2015}, we find that the \lya\ nebulae incidence rate ($\approx 20\%$) is lower than what is typically obtained in dedicated searches ($50-70\%$; e.g. \citealt{roche2014,arrigoni-battaia2016,arrigoni-battaia2019,farina2017,farina2019,zheng2019b,osullivan2020}), although we are limited by small-number statistics.

\begin{deluxetable*}{ccccccccc}
\tabletypesize{\normalsize}
\tablecaption{Survey results. The columns correspond to the QSO sightline identifier (1), DLA subsample (2), redshift of the DLA (3),  \HI\ column density of the DLA (4), metallicity of the DLA (5), redshift of the emission (\lya; 6), \lya\ line flux ($S/N_{Ly\alpha} \geq 5$; 7), \lya\ luminosity (8), and impact parameter of the galaxy (\lya; 9).}
\setlength{\tabcolsep}{0.037in}
\tablehead{
\colhead{QSO} & \colhead{Sample} & \colhead{$z_{\rm abs}$} & \colhead{$\log{N_{H\textsc{i}}}$} & \colhead{[M/H]} & \colhead{$z_{\rm em}$} & \colhead{F$_{\rm Ly\alpha}$} & \colhead{L$_{\rm Ly\alpha}$} & \colhead{b} \cr \colhead{} & \colhead{} & \colhead{} & \colhead{[$\log{\mbox{cm}^{-2}}$]} & \colhead{} & \colhead{} & \colhead{[$10^{-17}$ erg s$^{-1}$ cm$^{-2}$]} & \colhead{[$10^{42}$ erg s$^{-1}$]} & \colhead{[kpc]} 
}
\startdata 
B0458-020 & control & 2.0396 & 21.65 $\pm$ 0.1 & -1.12 $\pm$ 0.1 & 2.041 & $6.4 \pm 3$ (6.3$\sigma$) & $2 \pm 1$ & $0 \pm 6$ \\
J2206-1958a$^{\dagger}$ & control/CO(\xmark) & 1.9200 & 20.65 $\pm$ 0.1 & -0.60 $\pm$ 0.1 & 1.923 & $14 \pm 6$ (6.5$\sigma$) & $4 \pm 1.7$ & $12 \pm 6$ \\
J2222-0946 & control/CO(\xmark) & 2.3543 & 20.55 $\pm$ 0.2 & -0.53 $\pm$ 0.2 & 2.357 & $13 \pm 5$ (19.3$\sigma$) & $6 \pm 2$ & $8 \pm 4$ \\
J1305+0924 & CO(\xmark) & 2.0184 & 20.4 $\pm$ 0.2 & -0.16 $\pm$ 0.2 & -- & $<2.2$  &  $<0.7$ & --\\
B1228-113 & CO(\cmark) & 2.1928 & 20.6 $\pm$ 0.1 & -0.22 $\pm$ 0.1 & -- & $<3.7$  &  $<1.4$ & -- \\
B0201+365  & CO(\cmark) & 2.4628 & 20.4 $\pm$ 0.1  & -0.24 $\pm$ 0.1 & -- & $<1.5$  &  $<0.8$ & -- \\
B0551-366 & CO(\cmark) & 1.9622 & 20.5 $\pm$ 0.1  & -0.27 $\pm$ 0.2 & 1.963 & $12 \pm 5$ (9.2$\sigma$) &  $3.4\pm 1.4$ & $53 \pm 4$ \\
 &  & &  &  & 1.957 & $17 \pm 6$ (6.4$\sigma$) &  $5 \pm 1.8$ & $70 \pm 4$ \\
J2225+0527 & CO(\cmark) & 2.1310 & 20.7 $\pm$ 0.1 & -0.09 $\pm$ 0.1 & -- & $<2.5$  &  $<0.9$ & -- \\
J1709+3258 & CO($\sim$) & 1.8300 & 20.95 $\pm$ 0.2  & -0.28 $\pm$ 0.2 & --  & $<60$  &  $<15$ & -- \\
B1230-101 & CO($\sim$) & 1.9314 & 20.5 $\pm$ 0.1  & -0.18 $\pm$ 0.1 & -- & $<14$ &  $<3.9$  & -- \\
J0453-1305$^{*}$ & blind & 2.0666 & 20.5 $\pm$ 0.1  & -1.39 $\pm$ 0.1 & 2.067$^{*}$  & $4.2 \pm 3^{*}$ (4$\sigma$) & $1.4 \pm 1^{*}$ & $6 \pm 6^{*}$ \\
J2206-1958b$^{\dagger}$ & blind & 2.0762 & 20.4 $\pm$ 0.1  & -2.26 $\pm$ 0.1 & -- & $<2$ & $<0.7$  & -- \\
Q1755+578 & blind & 1.9692 & 21.4 $\pm$ 0.2 & -0.18 $\pm$ 0.2 & --  & $<4.8$ &  $<1.4$ & -- \\
J1013+5615 & blind & 2.2831 & 20.7 $\pm$ 0.2 & -0.07 $\pm$ 0.2 & -- & $<1.3$ & $<0.6$ & -- \\
\enddata
\tablecomments{The symbols in the second column denote significant detections (\cmark), tentative detections ($\sim$), and nondetections (\xmark) in the CO imaging. There are two DLAs along the sightline of J2206-1958 ($\dagger$). The DLA along B0551-366 has two detections in \lya. The DLA along J0453-1305 is a tentative \lya\ detection ($*$). The uncertainty on the \lya\ line flux includes the error on the absolute flux calibration, and thus it should not be used to estimate the significance of the line. The signal-to-noise ratios of the individual detections are listed in parenthesis, next to the \lya\ flux measurements.}
\label{table:results}
\vspace{0em}
\end{deluxetable*}

\begin{figure*}
\centering
\includegraphics[width=7.1in]{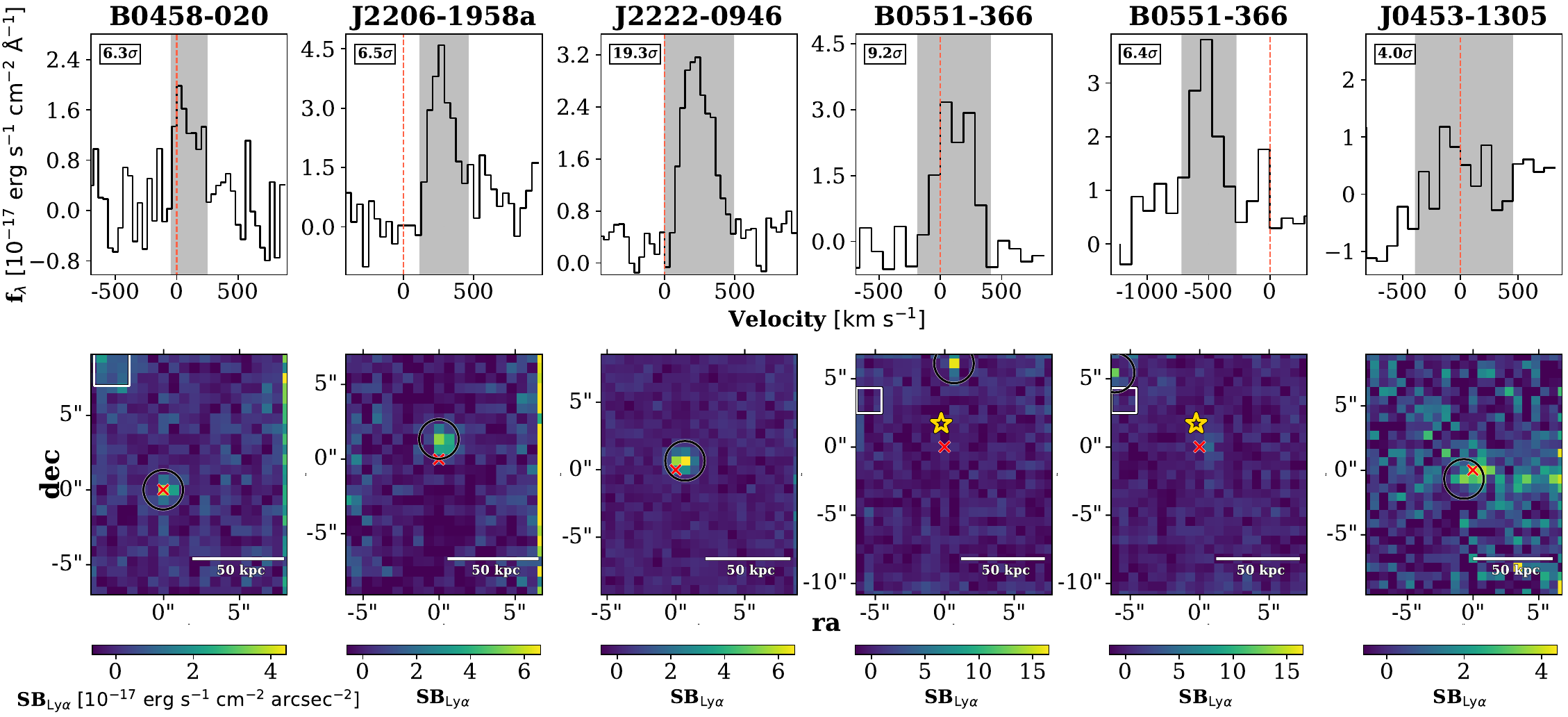}
\caption{Visualization of our line search for the 6 \lya\ detections. The six columns correspond to the six \lya\ emission lines. \textbf{Top:} spatially integrated \lya\ emission after subtraction of the QSO continuum. The red dashed line shows the \lya\ absorption wavelength. The Ly$\alpha$ emitters along B0458-020, J2206-1958, and J2222-0946 are part of our control sample, i.e., they were originally detected in searches for Ly$\alpha$ emitters around DLAs at low impact parameters. Two more \lya\ emitters at $b\approx 53$~kpc and $b\approx 70$~kpc away from DLA B0551-366 were also found. This DLA also has a CO-emitter at $b\approx 15$~kpc. The field showing a tentative detection --- J0453-1305 --- belongs in the blind sample, i.e., it has not been a target in any CO or \lya\ searches. \textbf{Bottom:} integrated flux of the spectrum in the KCWI cubes centered at the redshifted \lya\ absorption and within a window of width equal to the emission line. The symbols show the position of the QSO (red cross), \lya\ emission (circles), CO-emitter (yellow star), and interlopers (white squares).
}
\label{fig:cubes1}
\end{figure*}

\begin{figure*}
\centering
\includegraphics[width=7.1in]{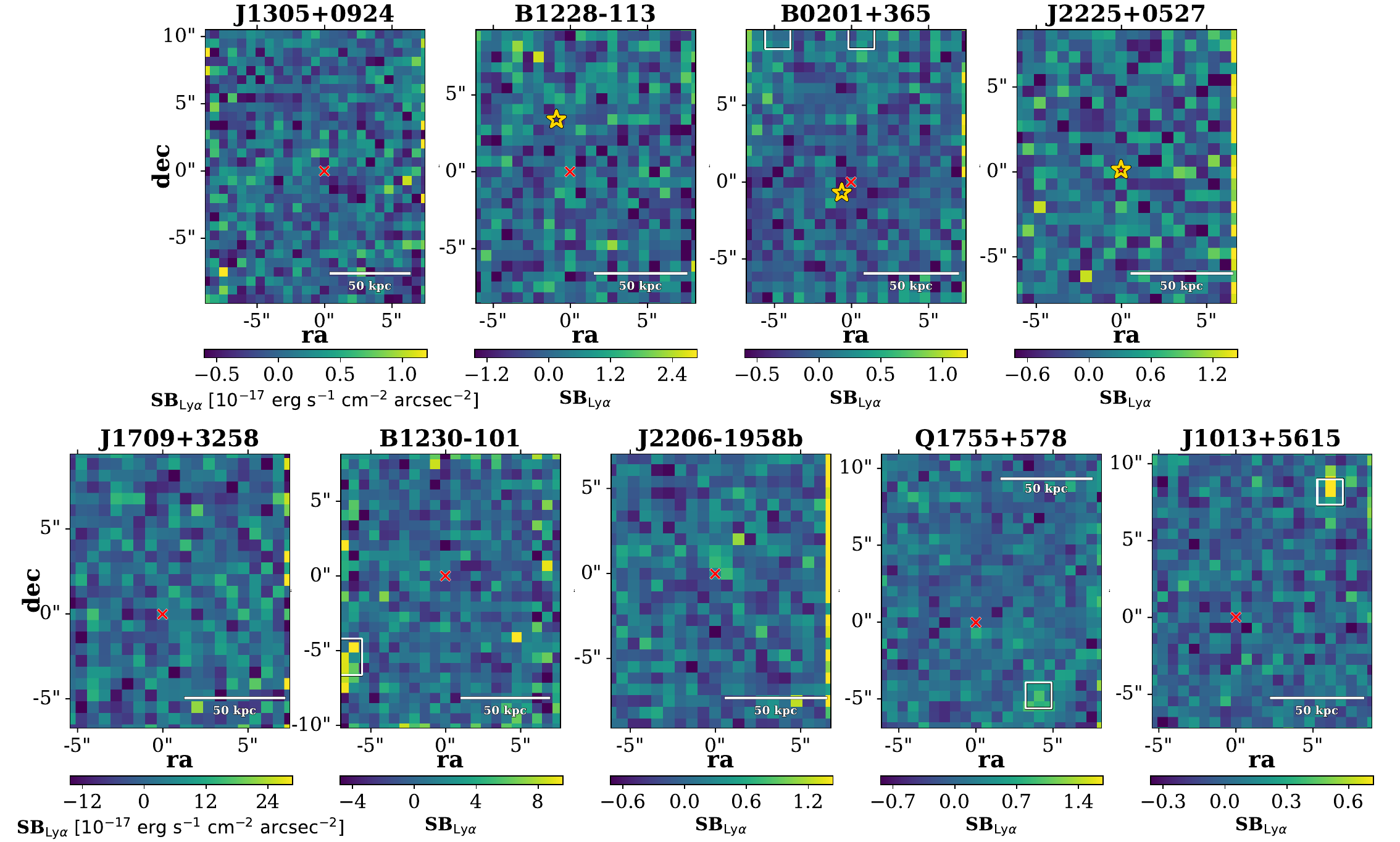}
\caption{KCWI cubes for all the \lya\ nondetections. Shown is the integrated flux of the spectrum in the KCWI cubes centered at the redshifted \lya\ absorption and within a window of width 350~km~s$^{-1}$. The position of the QSOs are shown with a red cross and the location of the CO-emitters is shown with yellow stars. Some fields show significant interloper emission that is spatially offset from the QSO (white squares).
}
\label{fig:cubes2}
\end{figure*}

\begin{figure*}
\centering
\includegraphics[width=7.1in]{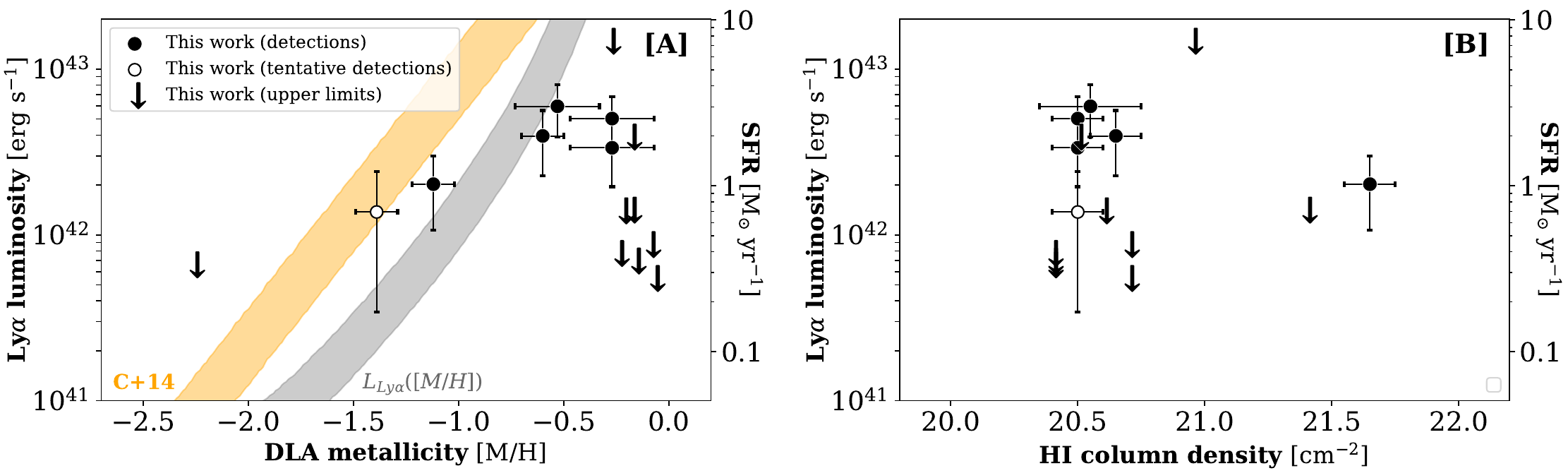}
\caption{The dependence of the \lya\ luminosity of the galaxies in the fields of DLAs at $z \approx 2$ on [A]~the DLA metallicity and [B]~the DLA \HI\ column density. Detections of \lya\ emission with $\geq 5\sigma$ significance are plotted as black circles, tentative ($4-5\sigma$) detections  as open circles, and \lya\ nondetections (i.e., $5\sigma$ upper limits to the \lya\ line luminosity) as downward-pointing arrows. The gray shaded region in panel~[A] indicates the expected \lya\ luminosity as a function of galaxy metallicity, which was obtained by combining the star-forming main sequence relation at $z \approx 2$ \citep{whitaker2014} with the observed stellar mass--metallicity relation for emission-selected galaxies at $z\approx 2$ 
\citep{erb2006}. The orange shaded region in this panel shows the expected Ly$\alpha$ luminosity as a function of DLA metallicity by combining the same star-forming main sequence relation at $z \approx 2$ with the mass-metallicity relation for \HI-selected galaxies at $z \approx 2$ (\citealt{christensen2014}). While a few of the \lya\ luminosities measured in our survey lie on or close to the grey band at intermediate metallicities, it is clear that  the upper limits to the \lya\ luminosity for the highest-metallicity DLAs are more than an order of magnitude below this band.
\label{fig:results_lya}}
\end{figure*}

\section{Results}
\label{5}

We detected 5 statistically significant Ly$\alpha$ emission lines in the fields of 4 of the 14 DLAs in our sample --- towards B0458-020 ($z \approx 2.041$), J2206-1958 ($z \approx 1.923$), J2222-0946 ($z \approx 2.357$), and B0551-366 ($z \approx 1.963$ and $z \approx 1.957$). The spatially integrated \lya\ emission spectra and the velocity-integrated \lya\ images for these detections are plotted in Figure~\ref{fig:cubes1}. Three of these DLAs belong to our \lya\ control sample, i.e., these three \lya\ emitters had been earlier identified via slit spectroscopy \citep{moller2002,moller2004,fynbo2010}. We also note that two of these control DLAs, towards J2222-0946 and J2206-1958, are part of the CO sample. No CO emission was detected by \citet{kanekar2020}. 

The fourth and fifth \lya\ detections are towards B0551-366. This DLA is part of the CO sample, featuring a CO detection. The \lya\ emitters are at impact parameters of $53$~kpc and $70$~kpc (see Figure~\ref{fig:cubes1}), which are much larger than the impact parameter of the CO emitter in this system ($15$~kpc). Remarkably, the \lya\ redshift of the galaxy that is $\approx 53$~kpc away from the DLA sightline is in excellent agreement with the DLA redshift ($\lesssim 100$~km~s$^{-1}$; Figure \ref{fig:cubes1}).

None of the 6 DLAs with significant or tentative CO detections show any evidence for \lya\ emission within $50$~kpc of the DLA (see Figures \ref{fig:cubes1} and \ref{fig:cubes2}). In passing, we note that the tentative CO detection at $z = 1.83$ towards J1709+3258 is just outside the field-of-view of our KCWI coverage. Finally, of the 4 DLAs in the ``blind'' sample, we obtained a tentative ($S/N_{Ly\alpha}  \approx 4$) detection of \lya\ emission at $z \approx 2.067$ towards J0453-1305. The \lya\ emission spectrum and image for this tentative detection are shown in the last column of Figure~\ref{fig:cubes1}. Our detections and nondetections of \lya\ emission are summarized in Table~\ref{table:results}.

The \lya\ emitter associated with the $z \approx 2.0395$ DLA along the sightline to B0458-020 was first reported by \citet[][see also \citealp{krogager2017}]{moller2004}. The galaxy was found to have a \lya\ flux of $(6.4 \pm 1.3) \times 10^{-17}$~erg~s$^{-1}$~cm$^{-2}$ at an impact parameter of $2.7 \pm 0.3$~kpc to the DLA sightline \citep{krogager2017}. For the $z \approx 2.3543$ DLA along J2222-0946, the \lya\ emitter was characterized in \citet{fynbo2010} and \citet{krogager2013}, with a \lya\ flux of $(14.3 \pm 0.3) \times 10^{-17}$~erg~s$^{-1}$~cm$^{-2}$ at an impact parameter of $6.3 \pm 0.3$~kpc \citep{krogager2017}. Finally, the galaxy associated with the $z \approx 1.9200$ DLA towards J2206-1958 was found to have a \lya\ flux of $(2.6 \pm 0.3) \times 10^{-16}$~erg~s$^{-1}$~cm$^{-2}$ at an impact parameter of $\approx 5.8$~kpc by \citet{moller2002}. In all three cases, 
our measurements for the \lya\ fluxes and impact parameters of the \lya\ emitters  (see Table~\ref{table:results}) are consistent (within $\approx 1.5\sigma$ significance) with the above measurements from the literature.

For the new \lya\ detections, the \lya\ emitters along the sightline towards B0551-366 have fluxes of $(12 \pm 5) \times 10^{-17}$~erg~s$^{-1}$~cm$^{-2}$ for the source at an impact parameter of $\approx 53$~kpc, and of $(17 \pm 6) \times 10^{-17}$~erg~s$^{-1}$~cm$^{-2}$ for the source at $b\approx 70$~kpc. For the tentative \lya\ detection associated with the $z \approx 2.0666$ DLA towards J0453-1305, we obtain a line flux of $(4.2 \pm 3) \times 10^{-17}$~erg~s$^{-1}$~cm$^{-2}$ at an impact parameter of $6 \pm 6$~kpc. We obtain $S/N_{Ly\alpha} \approx 4$ for this emitter, which was determined via simulations of the data cubes. Note that the error on the flux is dominated by the error in the flux scale.

Figure \ref{fig:results_lya} shows the results of our \lya\ searches from galaxies associated with DLAs at $z \approx 2$, with \lya\ luminosity plotted as a function of [A]~DLA metallicity and [B]~\HI\ column density. \lya\ detections are shown as filled black circles and \lya\ nondetections (and $5\sigma$ upper limits on the \lya\ luminosity) as downward-pointing arrows. The gray shaded region shows the expected \lya\ luminosity for a given galaxy metallicity, assuming the observed stellar mass--metallicity relation for star-forming galaxies at $z\approx 2$ \citep{erb2006} and the galaxy main-sequence relation at $z \approx 2$ \citep[with an assumed spread of $0.35$~dex;][]{whitaker2014,merida2023}. After using the H$\alpha$ SFR calibration of \citet{kennicutt2012}, a dust-free \lya\ to H$\alpha$ luminosity ratio of 8.7, and a \citep{chabrier2003} initial mass function \citep[e.g.][]{matthee2016,oyarzun2017,sobral2019}, the above relations were used to determine the expected H$\alpha$ line luminosity as a function of galaxy metallicity. Similarly, also shown in Figure~\ref{fig:results_lya} (orange) is the expected \lya\ luminosity if one takes the mass-metallicity relation of DLA absorption-selected galaxies with a scatter of 0.38 dex (\citealt{moller2013,christensen2014}).  

It is clear from Figure~\ref{fig:results_lya}[A] that most of our \lya\ detections arise in the fields of DLAs with relatively low absorption metallicities. Indeed, only one of the nine DLAs with [M/H]~$> -0.3$ shows a detection of \lya\ emission, while four of the five DLAs with [M/H]~$\leq -0.5$ show either clear or tentative detections of \lya\ emission. Further, the figure shows that many of the \lya\ detections of our survey lie close to the grey shaded region, i.e. are in reasonable agreement with the expected \lya\ line luminosity, despite the strong assumptions of (1)~a dust-free \lya\ to H$\alpha$ luminosity ratio and (2)~absorption metallicity equal to emission metallicity. This is seen to be the case for all DLAs with [M/H]~$\lesssim -0.5$. However, for the higher-metallicity DLAs, with [M/H]~$> -0.3$, most of the upper limits on the \lya\ line luminosity are in clear conflict with the expected line luminosity, typically lying more than an order of magnitude below the expected values. The most likely cause of this discrepancy is significant dust extinction of the \lya\ line in the galaxies in the fields of the highest-metallicity DLAs at $z \approx 2$ (more in Section \ref{6}).

While no clear trend is apparent in Figure~\ref{fig:results_lya}[B], which plots the \lya\ line luminosity versus DLA \HI\ column density, we cannot yet conclude that \lya\ luminosity is not correlated with \NHI\ for DLAs at $z \approx 2$. Apparent in this figure is that the coverage toward \NHI$> 10^{21}$~cm$^{-2}$ in current surveys is limited. While the survey by \citet{krogager2017} included some high \NHI\ DLAs, the associated galaxies are biased toward low impact parameters due to the slit-based nature of their search. Therefore, more data is needed to determine how \NHI\ and \lya\ luminosity are related at these redshifts.

\begin{figure*}
\centering
\includegraphics[width=7.1in]{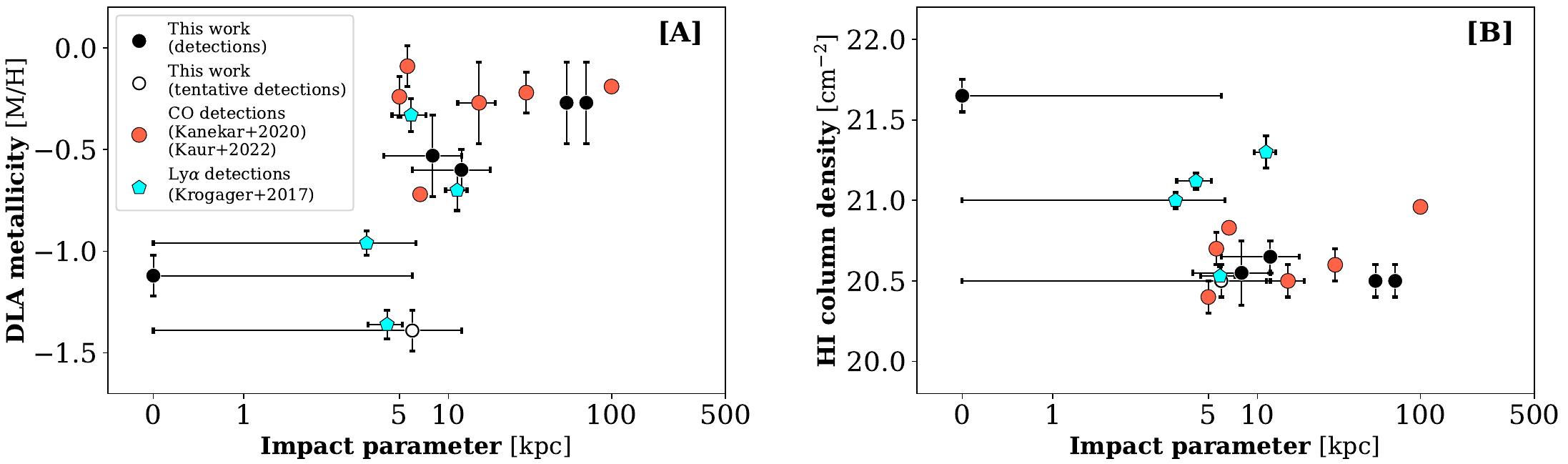}
\caption{[A]~DLA metallicity and [B]~DLA \HI\ column density plotted against galaxy impact parameter for DLA galaxies at $z\approx 2$. As in Figure \ref{fig:results_lya}, plotted are detections (black circles) and tentative detections (open circles). Also included are the slit spectroscopy measurements from \citet{krogager2017} and the CO emission measurements from \citet{kanekar2020} and \citet{kaur2022b,kaur2022a}.}
\label{fig:results_lya2}
\end{figure*}

The two panels of Figure~\ref{fig:results_lya2} show [A]~the DLA metallicity [M/H], and [B]~the DLA \HI\ column density, plotted against galaxy impact parameter for our six detections of \lya\ emission at $z \approx 2$. Besides these \lya\ detections, we have included the 6 DLA galaxies identified in ALMA and NOEMA CO searches at $z \approx 1.8-2.6$ \citep{kanekar2020,kaur2022b} and the five \lya\ detections at similar redshifts obtained via slit spectroscopy in the literature \citep{krogager2017}.

Figure~\ref{fig:results_lya2}[A] shows that there are galaxies over a wide range of impact parameters ($\approx 6-100$~kpc) in the fields of high-metallicity ([M/H]~$\gtrsim -0.7$) DLAs. This suggests that high-metallicity DLAs at $z \approx 2$ may arise from both galaxy disks and extended gas in the environment of massive galaxies, with the circumgalactic medium (CGM) being enriched due to galactic outflows. Conversely, the DLA galaxies associated with low metallicity ([M/H]~$\lesssim -1$) DLAs are seen to have low impact parameters ($b\lesssim 10$~kpc; although we note that there are only four galaxies in this category). Figure~\ref{fig:results_lya2}[B] plots the \HI\ column density against impact parameter for the above sample of \lya-detected and CO-detected galaxies. It is clear that similar \HI\ column densities, $\approx 10^{20.5} - 10^{21}$~cm$^{-2}$ are found in DLAs over a wide range of galaxy impact parameters, $\approx 5-100$~kpc.

\begin{figure*}
\centering
\includegraphics[width=7.1in]{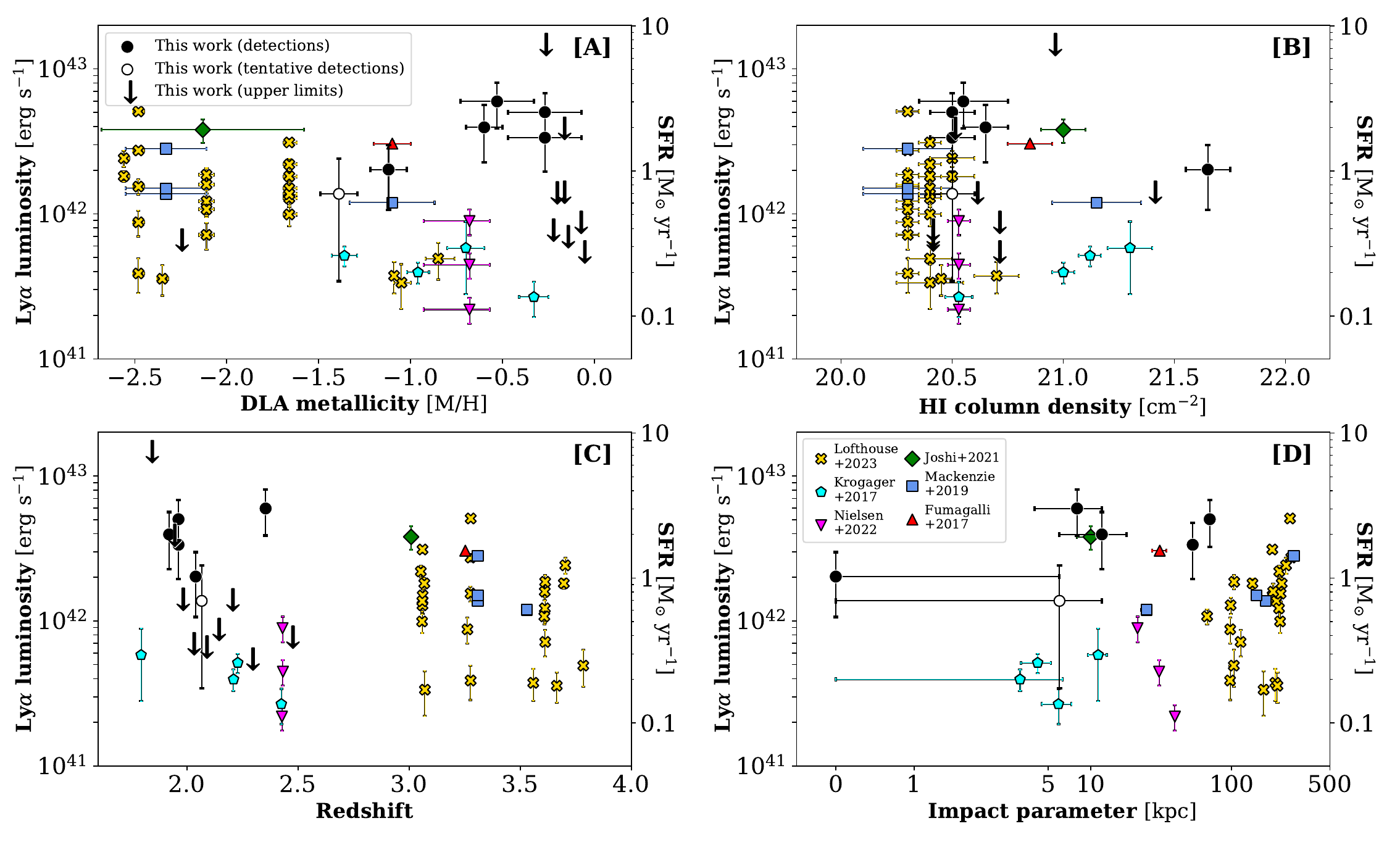}
\caption{The results of our survey in the context of searches for \lya\ emission from DLA fields at $z\gtrsim2$ in the literature. The panels show the Ly$\alpha$ luminosity as a function of [A]~DLA metallicity, [B]~\HI\ column density, [C]~redshift, and [D]~impact parameter. The results of our survey are shown as filled (detections) and open (tentative detections) circles, with upper limits to the \lya\ luminosity indicated by the downward-pointing arrows. Plotted from the literature are the measurements of \citet{krogager2017}, \citet{fumagalli2017}, \citet{mackenzie2019}, \citet{joshi2021}, \citet{nielsen2022}, and \citet{lofthouse2023}. The corresponding symbols are indicated in [D].
}
\label{fig:results_literature}
\end{figure*}

Finally, Figure \ref{fig:results_literature} compares our results to those obtained from IFU- and slit-based \lya\ searches in the fields of high-redshift DLAs in the literature. Included in this figure are the \lya\ detections and nondetections from our work, four slit-based detections around four DLAs at $z \approx 2$ \citep{krogager2017}, 5 \lya\ emitters around 3 DLAs at $z\gtrsim 3$ \citep{fumagalli2017,mackenzie2019}, a \lya\ emitter at $z\approx 3$  \citep{joshi2021}, 3 \lya\ emitters around a $z\approx 2.4$ DLA  \citep{nielsen2022}, and 23 \lya\ emitters around 9 DLAs at $z \approx 3-3.8$ \citep{lofthouse2023}. We emphasize that the \lya\ detections shown in this figure span a wide range of impact parameters, and that these previous observations have shown that DLAs can arise from galaxy groups \citep[e.g.,][]{fynbo2018}. It is therefore likely that some of the galaxies at large impact parameter are companion galaxies of the galaxy that gives rise to the DLA absorption \citep[e.g.][]{mackenzie2019,lofthouse2023}. This is consistent with the results from hydrodynamical simulations \citep[e.g.,][]{rahmati-schaye2014,rhodin2019}.

The different panels in Figure \ref{fig:results_literature} show how the \lya\ luminosity depends on [A]~DLA metallicity, [B]~\HI\ column density, [C]~redshift, and [D]~impact parameter. We note that the DLAs with searches for \lya\ emission at $z \gtrsim 3$ typically have low metallicities, [M/H]~$\lesssim -1$, while most of our KCWI searches are in the fields of high-metallicity DLAs. The luminosity of our upper limits decreases with redshift, reflecting how the spectral signal-to-noise decreases substantially for $z<2$ as a result of the degradation in sensitivity of KCWI blueward of 3700 \AA. 

Figure~\ref{fig:results_lya2}[D] might hint that DLAs at $z \gtrsim 3$ are more often associated with galaxy groups (with $\geq 3$~galaxies) than DLAs at $z \approx 2$. Four out of 13 DLAs at $z \gtrsim 3$ feature at least 3 \lya\ emitters in their fields \citep{mackenzie2019,lofthouse2023}, while only a single DLA at $z \approx 2$ (out of 15) has $\geq 3$ associated \lya\ emitters \citep{nielsen2022}. However, this apparent difference could be driven by variations in survey design. There is a significant difference between the field-of-views of KCWI ($\approx 170~{\rm kpc} \, \times 277$~kpc) and VLT-MUSE ($\approx 450 \, {\rm kpc} \times 450 \, {\rm kpc}$), the instrument used by the surveys of \citet{mackenzie2019} and \citet{lofthouse2023}. This difference becomes even greater when we consider that our effective field-of-view is $\approx 170~{\rm kpc} \, \times 130$~kpc (Section \ref{3}), which can be even smaller if the QSO is not perfectly centered (Figures \ref{fig:cubes1} and \ref{fig:cubes2}). Difference in sensitivities between the surveys could also play a role, with \citet{mackenzie2019} and \citet{lofthouse2023} achieving lower \lya\ luminosities than our analysis (Figure \ref{fig:results_literature}). Thus, it is clear that searches with wider fields of view and higher sensitivities at $ z \approx 2$ are needed to test for possible redshift evolution in DLA environments.

\section{Discussion}
\label{6}

Our KCWI survey has yielded new detections of \lya\ emission along the DLA sightline towards B0551-366, a tentative detection from the DLA field towards J0453-1305, and recoveries of all three known \lya\ detections from the control sample. Restricting to the new (i.e. non-control) searches, we have obtained definite \lya\ detections along only a single DLA sightline out of 11, with nine of the target DLAs having high metallicities ([M/H]~$> -0.3$).
We also do not detect \lya\ emission from any of the 4 CO-emitting galaxies in the sample, although one of these sightlines (B0551-366) shows \lya\ emission from two other galaxies within $\pm 500$~km~s$^{-1}$ of the CO and DLA redshifts. We discuss in this section the implications of these results, in conjunction with those from surveys from the literature, to gain insights into the nature of the galaxies associated with high-metallicity DLAs at $z \approx 2$.  

\subsection{The galaxies associated with high-metallicity DLAs at $z\approx2$ are not typical \lya\ emitters}
\label{6.1}

The left panel of Figure \ref{fig:results_lya} shows that the \lya\ luminosity of galaxies in the field of high-metallicity ([M/H]~$\geq -0.3$) DLAs at $z\approx 2$ lies well below the predicted \lya\ luminosity of emission-selected galaxies at similar redshifts. We emphasize that the predicted \lya\ luminosity has been obtained with the assumption of a dust-free \lya-to-H$\alpha$ ratio. There are two possibilities to explain this discrepancy: (1)~the galaxies associated with high-metallicity DLAs have low stellar masses, and hence intrinsic \lya\ luminosities below our detection threshold, or (2)~the assumption of a dust-free \lya-to-H$\alpha$ ratio breaks down, with the galaxies associated in high-metallicity DLA fields being typically massive, dusty galaxies with high production, high absorption of \lya\ photons. In this section, we consider the first possibility by turning to two different methods of estimating the luminosities of low mass (M$_*\approx  10^8 \, \rm M_{\odot}$) galaxies. As shown in \citet{oyarzun2016,oyarzun2017}, galaxies of this stellar mass have \lya\ equivalent widths ($50-200$\AA; \citealt{charlot1993}) and escape fractions ($\approx 1$; \citealt{laursen2009}) that are typical of dust-free interstellar media.

In the first approach, we will assume that any undetected galaxies belong to the star-formation main sequence at $z \approx 2$. The slope and scatter of the main sequence at $z\approx 2$ has been measured down to a stellar mass of M$_*\approx 10^8 \, \rm M_{\odot}$ by \citet{merida2023}. At this mass, the SFR distribution has an average of log(SFR/M$_{\odot}$ yr$^{-1})\sim 0$ and a scatter of $\Delta$log(SFR/M$_{\odot}$ yr$^{-1})\sim 0.35$ dex. Following the approach of Section \ref{5}, this SFR can be converted into a \lya\ luminosity L$_{Ly\alpha}$ by assuming a dust-free \lya-to-H$\alpha$ ratio and the H$\alpha$ SFR calibration for a Chabrier IMF. In the resulting L$_{Ly\alpha}$ distribution, the probability that a galaxy has a \lya\ luminosity lower than our upper limits is  $p\approx 20\%$ (on average) for the 8 high-metallicity DLAs with tight upper limits on the \lya\ luminosity (i.e. J1305+0924, B1228-113, B0201+365, J2225+0527, B1230-101, Q1755+578, and J1013+5615). Thus, the probability that all of these DLAs feature typical \lya\ emitters with M$_*\approx 10^8 \, \rm M_{\odot}$ within $\approx 50$~kpc of the QSO sightline is $p\lesssim 10^{-5}$.

Alternatively, we can turn to the rest-frame \lya\ equivalent width (EW$_{Ly\alpha}$) instead of the \lya\ luminosity. This quantity is defined as
\begin{equation}
\mbox{EW}_{Ly\alpha} = \frac{\mbox{F}_{Ly\alpha}}{\mbox{f}_{\lambda}} \frac{1}{(1+z)},
\end{equation}
where F$_{Ly\alpha}$ is the \lya\ line flux and $f_{\lambda}$ is the rest-frame flux density of the galaxy in the near-UV. We note that the EW$_{Ly\alpha}$ is a convenient metric because the EW$_{Ly\alpha}$ distribution of high-redshift galaxies has been thoroughly quantified in the literature (e.g. \citealt{gronwall2007,treu2012,jiang2013a}).

For low-mass galaxies with M$_*\approx 10^8 \, \rm M_{\odot}$ and log(SFR/M$_{\odot}$ yr$^{-1})\sim 0$, the rest-frame near-UV flux densities are typically in the range $f_{\lambda} \sim (1-10) \times 10^{-19}$ erg s$^{-1}$ cm$^{-2}$ \AA$^{-1}$ (e.g. \citealt{skelton2014}). With $f_{\lambda}$ in hand and the upper limits on the \lya\ line flux density, we obtain typical upper limits to the rest-frame equivalent width of  EW$_{Ly\alpha} \lesssim 30$\AA\ within $\approx 50$~kpc of the 8 high-metallicity DLAs in our sample.

The expected \lya\ detection rate of our observations can now be estimated from the known EW$_{Ly\alpha}$ distribution of galaxies at $z \approx 2$. A number of authors have concluded that the EW$_{Ly\alpha}$ distribution at this redshift is well fitted by an exponential profile with a scale length of $EW_0 \approx 50$\AA\ \citep[e.g.][]{nilsson2009,guaita2010,mawatari2012,ciardullo2014}. Knowing that the EW$_{Ly\alpha}$ distribution accounts for $\approx 80\%$ of the galaxy population at M$_*\lesssim 10^9 \, \rm M_{\odot}$ ($z\sim 4$; \citealt{oyarzun2016}), integration of this probability distribution from EW$_{Ly\alpha} = 0$ up to our EW$_{Ly\alpha}$ limits yields $\approx 80\%$ of the probability that a M$_*\approx 10^8 \, \rm M_{\odot}$ galaxy was not detected by our survey. This results in total probabilities of $\approx 40\%$ for each of our 8 nondetections, yielding a probability of $p \approx 2 \times 10^{-4}$ that all of these 8 high-metallicity DLAs have low stellar mass galaxies within $\approx 50$~kpc of the QSO sightline.

These results indicate that the majority of the galaxies associated with high-metallicity DLAs at $z\approx 2$ are not unobscured galaxies with low stellar masses, i.e., M$_*\approx  10^8 \, \rm M_{\odot}$. In fact, characterization of the galaxies associated with high-metallicity DLAs at $z\approx 2$ has yielded stellar masses exceeding $10^{10}$~M$_{\odot}$ (e.g. \citealt{fynbo2013}).

\newpage

\subsection{The galaxies associated with high-metallicity DLAs at $z\approx2$ are probably massive and dusty}

\label{6.2}

A galaxy can also remain undetected in \lya\ emission if it has a low \lya\ escape fraction. This would typically arise because of \lya\ photon absorption by dust grains, as indicated by several radiative transfer simulations \citep[e.g.][]{verhamme2008,laursen2009}. This has been argued to be the reason for why high-redshift galaxies with redder UV slopes show lower \lya\ emission equivalent widths \citep[e.g.][]{blanc2011,hayes2011,atek2014,oyarzun2017}. At the same time, the escape of \lya\ photons is affected by the scattering of radiation by neutral gas, such that the structure and kinematics of the interstellar and circum-galactic media of galaxies can shape the \lya\ line profile (e.g. \citealt{verhamme2006}).

The dependence of the \lya\ escape fraction on the neutral gas covering fraction and dust extinction has implications for the use of \lya\ as a galaxy tracer. Star-forming galaxies with higher neutral gas masses and dust extinctions tend to have higher stellar masses \citep[e.g.][]{reddy2006,finlator2007,tacconi2020}, implying that \lya\ emission might not be a good tracer of the high stellar mass end of the star-forming galaxy population. In agreement with this picture, the equivalent width of the \lya\ emission line in high-redshift galaxies has been found to anti-correlate with the stellar mass \citep[e.g.][]{oyarzun2016}.

Instead, massive galaxies tend to be bright in mm or sub-mm emission lines (e.g. CO or [C\,{\sc ii}]~158$\mu$m), whose luminosity correlates with the total molecular gas mass (i.e. with the stellar mass). Thus, it is not surprising that the CO rotational lines have been particularly useful in identifying the galaxies associated with the highest-metallicity DLAs at $z \approx 2$ \citep[]{kanekar2020,kaur2022a}. \citet{kaur2022a} find that the detection rate of CO emission is strongly dependent on DLA metallicity, with a CO detection rate of $\approx 50\%$ for [M/H]$>-0.3$, and a far lower CO detection rate at lower metallicities.

The suggestion that \lya\ and CO emission are efficient at identifying the galaxies associated with DLAs of different metallicities is supported by Figures~\ref{fig:results_lya2}[A] and \ref{fig:results_literature}. Galaxies detected through \lya\ emission tend to be brighter for DLAs with [M/H]~$\lesssim-1.0$. This is likely to arise due to dust obscuration effects, with higher-metallicity galaxies also having high dust contents that impede the escape of \lya\ photons. Conversely, the requirements of both a high molecular gas mass and a low CO-to-H$_2$ conversion factor imply that searches for CO emission favour the identification of the high-mass galaxies associated with high-metallicity DLAs. In line with the detection of dusty galaxies in association with high-metallicity DLAs, it is plausible that the majority of these absorbers arise from massive, dusty galaxies that are faint in \lya\ emission \citep[e.g.][]{fynbo2013}.

In this context, it is interesting to consider why some of the highest-metallicity DLAs show no evidence of CO emission from associated galaxies \citep{kanekar2020,kaur2022a}. This is likely a result of the sensitivities of the current ALMA and NOEMA searches, which typically yield upper limits of $\approx 1-5 \times 10^{10} \, \rm M_\odot$ on the molecular gas masses of these galaxies \citep{kanekar2020,kaur2022a}. These upper limits are not necessarily constraining, and thus current CO nondetections do not rule out the presence of massive galaxies in the DLA fields.

Given that high-mass galaxies associated with high-metallicity DLAs may be missed in searches for both \lya\ and CO emission, turning to different diagnostics emerges as a valid strategy. The best probes to identify such galaxies are likely to be H$\alpha$, [O\,{\sc iii}]$\lambda$5007, and [C\,{\sc ii}] 158$\mu$m emission. The H$\alpha$ and [O\,{\sc iii}]$\lambda$5007 lines  are less affected by dust obscuration than \lya, and can be used to probe galaxies in the stellar mass range M$_*=10^{8}-10^{10} \, \rm M_\odot$ \citep[e.g.][]{peroux2011,peroux2012,jorgenson2014,wang2015}. While this line is difficult to access from the ground for galaxies at $z>2$, the James Webb Space Telescope should allow the H$\alpha$-based identification of large samples of high-mass DLA galaxies out to $z\sim 4$. 
Similarly, while the [C\,{\sc ii}]~158$\mu$m emission from galaxies at $z \approx 2$ lies at very high frequencies ($> 600$~GHz), it should be possible to use ALMA [C\,{\sc ii}]~158$\mu$m searches to identify such galaxies. 

\subsection{Implications for the nature of \HI-selected galaxies at high redshift}

Searches for the galaxies associated with DLAs at high redshift have made remarkable progress over the last few years, with more than 40~galaxies identified via \lya, CO, or [C\,{\sc ii}]~158$\mu$m searches \citep[e.g.][]{krogager2017,fumagalli2017,neeleman2017,neeleman2018,neeleman2019,mackenzie2019,kanekar2020,kaur2022a,lofthouse2023}. While we now have a large sample of \HI-selected galaxies, the wide range of DLA redshifts ($z\approx 2-4.5$), DLA metallicities ([M/H]~$-2.5 - 0$), and different selection techniques imply that it is not straightforward to draw conclusions about their nature. Here, we briefly summarize the current observational view of this population in order of increasing redshift.

First, the bulk of the \HI-selected galaxies identified in slit-based \lya\ searches at $z \approx 2$ lie at low impact parameters to the QSO sightline \citep[e.g.][]{fynbo2013,krogager2016,krogager2017,joshi2021}. This is not surprising, given that such searches are only sensitive towards galaxies at low  impact parameters ($\lesssim 15$~kpc). By design, these DLAs have high metallicities, i.e., between [M/H]~$\approx -1.5$ and [M/H]~$\approx-0.5$ \citep{krogager2017}. The fraction of undetected galaxies and their \HI\ metallicity distribution remains unclear.

At $z \approx 2$, our KCWI \lya\ survey has revealed only 2-3 new \HI-selected galaxies from a search in 11 DLA fields (excluding the control sample). Nine of our DLA targets have [M/H]~$> -0.3$, of which only one showed a detection of \lya\ emission. The KCWI survey is sensitive to galaxies with impact parameters $\lesssim 50$~kpc.

At $z \approx 2$, ALMA and NOEMA CO searches have yielded six definite detections of \HI-selected galaxies, all in the fields of DLAs with a metallicity of [M/H]~$> -0.3$. \citep{kanekar2020,kaur2022a}. Five of the CO detections are at impact parameters of $\lesssim 30$~kpc. The sixth system, towards Q0918+1636, has an impact parameter of $\approx 100$~kpc; however, there is a galaxy at lower impact parameter ($\approx 16$~kpc) to the QSO sightline that was identified in the optical \citep{fynbo2018}. While the field of  Q0918+1636 was not targeted by our survey, the DLA galaxy at an impact parameter of $\approx 16$~kpc has been extensively characterized by \citet{fynbo2011}. The nondetection of \lya\ emission from this galaxy down to fluxes of $\sim 5 \times 10^{-18}$~erg~s$^{-1}$~cm$^{-2}$ is indicative of dust supression, especially given the high luminosity of its [O\,{\sc ii}] and [O\,{\sc iii}] emission \citep{fynbo2011}.

At $z \approx 3-3.5$, \lya\ spectroscopy with VLT/MUSE has yielded $\approx 28$ \lya\ detections of galaxies in DLA fields \citep{fumagalli2017,mackenzie2019,lofthouse2023}. These DLAs have metallicities between [M/H]~$\approx -2.5$ and [M/H]~$\approx -1$. Almost all of the \lya\ detections are at high impact parameters ($\gtrsim 100$~kpc), and there are, in some cases, more than 5 galaxies identified in a single DLA field. Such galaxies would not have been detected either in the slit-based \lya\ surveys (because of the field of view) or in the present KCWI survey (because of the sensitivity and/or field of view).

Finally, searches in [C\,{\sc ii}]~158$\mu$m at $z \approx 3.8-4.5$ with ALMA  have identified 10 galaxies in the fields of DLAs with [M/H]$\geq -1.3$ \citep[e.g.][]{neeleman2017,neeleman2019,prochaska2019,kaur2021}. The impact parameters to the QSO sightline are large ($15-50$~kpc). Three of the DLA fields have more than 2 galaxies within $\approx 50$~kpc of the QSO sightline.

Regarding the sizes of \HI\ gas reservoirs, evidence indicates that they are quite extended, especially in low-redshift massive galaxies. Towards $z \approx 1.3$, \citet{chowdhury2022} have found that a spatial resolution of 90~kpc is needed to recover the total \HI~21\,cm emission of massive galaxies, indicating that the \HI\ sizes of these galaxies are $\gtrsim 50$~kpc at these redshifts. In the local Universe, the \HI\ size has been found to correlate with the \HI\ mass, with the diameter of \HI\ disks exceeding 40~kpc in galaxies with \HI\ masses $>10^{10}$~M$_{\odot}$ \citep[at an \HI\ surface density of $1 \rm M_\odot$/pc$^2$, i.e., similar to the DLA column density threshold; e.g.][]{broeils97,wang2016b}. Thus, both direct measurements of \HI~21\,cm emission at $z \approx 1.3$ and the \HI\ mass-size relation at $z \approx 0$ suggest that massive galaxies have large spatial extents in \HI.

\citet{chowdhury2022b} have concluded that \HI\ dominates the baryonic mass of the disks of galaxies, with the \HI\ mass exceeding the stellar mass and the molecular gas mass by factors of $\approx 4-5$. Drawing a comparison with the CO detections of \citet{kanekar2020} and \citet{kaur2022a}, the measurements by \citet{chowdhury2022b} would imply an \HI\ diameter of $\gtrsim 100$~kpc at $z \approx 2$, even if we assume that the \HI\ mass is only comparable to the molecular gas mass, ($\gtrsim 5 \times 10^{10} \, \rm M_\odot$). With the \HI-selected galaxies associated with high-metallicity ([M/H]~$\gtrsim -0.5$) DLAs at $z \approx 2$ typically found at relatively low impact parameters ($\lesssim 30$~kpc), it is plausible that high-metallicity DLAs at these redshifts arise from the disks of massive galaxies.

As concluded in Section \ref{6.2}, the low detection rate of these high-metallicity DLA fields in \lya\ is likely due to dust obscuration effects in massive galaxies. Instead, less affected by dust obscuration (and brighter in \lya) are expected to be galaxies of low-intermediate stellar masses (Section \ref{6.1}). Because of this, it is noteworthy how small the number of companion galaxies identified in \lya\ is at $z \approx 2$. Only one of the \HI-selected galaxies at low impact parameters (B00551-366) has been found to have companion galaxies within $\approx 50$~kpc. This is likely to be due to relatively small fields of view of the current \lya\ searches at $z \approx 2$.

The situation is somewhat different at $z \approx 3-3.5$, where \lya\ searches have mostly identified galaxies at very large impact parameters ($\gtrsim 100$~kpc) in the fields of DLAs with mostly low metallicities (between [M/H]~$\approx -2.5$ and [M/H]~$\approx -1$; \citealt{mackenzie2019,lofthouse2023}). If we assume that the low metallicity of the DLA is indication of low dust obscuration, galaxies at low impact parameters would have only been missed if they were intrinsically under-luminous. Under this assumption, DLAs at $z \approx 3-3.5$ with [M/H]~$\lesssim -1$ arise from either low stellar mass galaxies at low impact parameters or from massive, dusty galaxies at large impact parameters. The large number of \lya-emitting companions at distances of $\approx 100-200$~kpc in a number of the fields is interesting, and suggests that the DLAs probed by these studies arise mostly in galaxy groups.

Finally, the impact parameters of the \HI-selected galaxies identified in [C\,{\sc ii}]~158$\mu$m searches at $z \gtrsim 4$ are $\approx 15-50$~kpc \citep{neeleman2017,neeleman2019}. These [C\,{\sc ii}]~158$\mu$m emitters are all associated with DLAs that have a metallicities of [M/H]~$\geq -1.35$, and while multiple galaxies have been identified in some fields, these are all at similar impact parameters ($<50$~kpc). Together, the high \HI\ column densities ($\gtrsim 10^{21}$~cm$^{-2}$) and relatively large impact parameters hint that high-metallicity, high \NHI\ DLAs at $z \gtrsim 4$ likely arise from \HI\ clumps in the CGM of massive galaxies. This is consistent with numerical simulations that predict greater amounts of \HI\ in the CGM of galaxies at $z \gtrsim 4$ than at lower redshifts \citep[e.g.][]{stern2021}.

\section{Summary}
\label{7}

We used the integral field spectrograph KCWI on the Keck~II telescope to carry out a search for \lya\ emission in the fields of 14 DLAs at $z \approx 2$. Nine of the 14 targets have high metallicities ([M/H]~$> -0.3$). Seven of the 14 fields have been searched for CO emission with ALMA or NOEMA, with 4 confirmed detections. Finally, three of the 14 DLAs are known \lya\ emitters from the literature that were observed to quantify our \lya\ detection capability (i.e., the control sample).

We detected \lya\ emission with the expected strength from the three control-sample DLAs. For the remaining 11 targets, \lya\ emission was detected from two galaxies in the field of the $z \approx 1.9622$ DLA towards B0551-366 at impact parameters of $\approx 50-70$~kpc. Also in the field of B0551-366 is a massive CO-detected galaxy at an impact parameter of $\approx 15$~kpc from the QSO sightline. This indicates that the \lya\ emitters do not directly give rise to the DLA absorption. We find that the low \lya\ detection rate in the fields of high-metallicity DLAs is likely a result of \lya\ photon absorption by dust produced in massive and dusty galaxies.

We compared the results of our \lya\ searches in DLA fields at $z \approx 2$ with those of CO searches at $z \approx 2$, \lya\ searches at $z \approx 3-3.5$, and [C\,{\sc ii}]~158$\mu$m searches at $z \approx 4$. The impact parameters of the galaxies associated with high-metallicity DLAs at $z \approx 2$ are typically $\lesssim 30$~kpc. We argue that high-metallicity galaxies are likely to have a large \HI\ mass, and hence a large \HI\ spatial extent. High-metallicity DLAs at $z \approx 2$ are thus likely to arise in the \HI\ reservoirs of massive galaxies.  

Finally, we found that galaxies associated with high-metallicity DLAs ([M/H]~$> -0.3$) may remain unidentified in both CO searches (if they do not have extreme molecular gas masses) and \lya\ searches (due to high dust obscuration). Searches in the H$\alpha$, [O\,{\sc iii}]$\lambda$5007, and [C\,{\sc ii}]~158$\mu$m lines will likely be necessary to identify this population.

\medskip

We thank the reviewer for detailed comments on an earlier version of this manuscript. This material is based upon work supported by the National Science Foundation under grants AST-2107989 and AST-2107990. MR also acknowledges support from the STScI's Director's Discretionary Research Funding (grant ID D0101.90306). RAJ gratefully acknowledges support from the Nantucket Maria Mitchell Association. NK acknowledges the Department of Atomic Energy for funding support, under project 12-R\&D-TFR-5.02-0700. The data presented herein were obtained at the W. M. Keck Observatory, which is operated as a scientific partnership among the California Institute of Technology, the University of California and the National Aeronautics and Space Administration. The Observatory was made possible by the generous financial support of the W. M. Keck Foundation. The authors wish to recognize and acknowledge the very significant cultural role and reverence that the summit of Maunakea has always had within the indigenous Hawaiian community.  We are most fortunate to have the opportunity to conduct observations from this mountain.

\bibliography{sample631}{}

\begin{thebibliography}{}
\expandafter\ifx\csname natexlab\endcsname\relax\def\natexlab#1{#1}\fi
\providecommand{\url}[1]{\href{#1}{#1}}
\providecommand{\dodoi}[1]{doi:~\href{http://doi.org/#1}{\nolinkurl{#1}}}
\providecommand{\doeprint}[1]{\href{http://ascl.net/#1}{\nolinkurl{http://ascl.net/#1}}}
\providecommand{\doarXiv}[1]{\href{https://arxiv.org/abs/#1}{\nolinkurl{https://arxiv.org/abs/#1}}}

\bibitem[{{Alam} {et~al.}(2015){Alam}, {Albareti}, {Allende Prieto}, {Anders},
  {Anderson}, {Anderton}, {Andrews}, {Armengaud}, {Aubourg}, {Bailey}, {Basu},
  {Bautista}, {Beaton}, {Beers}, {Bender}, {Berlind}, {Beutler}, {Bhardwaj},
  {Bird}, {Bizyaev}, {Blake}, {Blanton}, {Blomqvist}, {Bochanski}, {Bolton},
  {Bovy}, {Shelden Bradley}, {Brandt}, {Brauer}, {Brinkmann}, {Brown},
  {Brownstein}, {Burden}, {Burtin}, {Busca}, {Cai}, {Capozzi}, {Carnero
  Rosell}, {Carr}, {Carrera}, {Chambers}, {Chaplin}, {Chen}, {Chiappini},
  {Chojnowski}, {Chuang}, {Clerc}, {Comparat}, {Covey}, {Croft}, {Cuesta},
  {Cunha}, {da Costa}, {Da Rio}, {Davenport}, {Dawson}, {De Lee}, {Delubac},
  {Deshpande}, {Dhital}, {Dutra-Ferreira}, {Dwelly}, {Ealet}, {Ebelke},
  {Edmondson}, {Eisenstein}, {Ellsworth}, {Elsworth}, {Epstein}, {Eracleous},
  {Escoffier}, {Esposito}, {Evans}, {Fan}, {Fern{\'a}ndez-Alvar}, {Feuillet},
  {Filiz Ak}, {Finley}, {Finoguenov}, {Flaherty}, {Fleming}, {Font-Ribera},
  {Foster}, {Frinchaboy}, {Galbraith-Frew}, {Garc{\'\i}a},
  {Garc{\'\i}a-Hern{\'a}ndez}, {Garc{\'\i}a P{\'e}rez}, {Gaulme}, {Ge},
  {G{\'e}nova-Santos}, {Georgakakis}, {Ghezzi}, {Gillespie}, {Girardi},
  {Goddard}, {Gontcho}, {Gonz{\'a}lez Hern{\'a}ndez}, {Grebel}, {Green},
  {Grieb}, {Grieves}, {Gunn}, {Guo}, {Harding}, {Hasselquist}, {Hawley},
  {Hayden}, {Hearty}, {Hekker}, {Ho}, {Hogg}, {Holley-Bockelmann}, {Holtzman},
  {Honscheid}, {Huber}, {Huehnerhoff}, {Ivans}, {Jiang}, {Johnson},
  {Kinemuchi}, {Kirkby}, {Kitaura}, {Klaene}, {Knapp}, {Kneib}, {Koenig},
  {Lam}, {Lan}, {Lang}, {Laurent}, {Le Goff}, {Leauthaud}, {Lee}, {Lee},
  {Licquia}, {Liu}, {Long}, {L{\'o}pez-Corredoira}, {Lorenzo-Oliveira},
  {Lucatello}, {Lundgren}, {Lupton}, {Mack}, {Mahadevan}, {Maia}, {Majewski},
  {Malanushenko}, {Malanushenko}, {Manchado}, {Manera}, {Mao}, {Maraston},
  {Marchwinski}, {Margala}, {Martell}, {Martig}, {Masters}, {Mathur},
  {McBride}, {McGehee}, {McGreer}, {McMahon}, {M{\'e}nard}, {Menzel},
  {Merloni}, {M{\'e}sz{\'a}ros}, {Miller}, {Miralda-Escud{\'e}}, {Miyatake},
  {Montero-Dorta}, {More}, {Morganson}, {Morice-Atkinson}, {Morrison},
  {Mosser}, {Muna}, {Myers}, {Nandra}, {Newman}, {Neyrinck}, {Nguyen},
  {Nichol}, {Nidever}, {Noterdaeme}, {Nuza}, {O'Connell}, {O'Connell},
  {O'Connell}, {Ogando}, {Olmstead}, {Oravetz}, {Oravetz}, {Osumi}, {Owen},
  {Padgett}, {Padmanabhan}, {Paegert}, {Palanque-Delabrouille}, {Pan},
  {Parejko}, {P{\^a}ris}, {Park}, {Pattarakijwanich}, {Pellejero-Ibanez},
  {Pepper}, {Percival}, {P{\'e}rez-Fournon}, {P{\textasciiacute}rez-Ra`fols},
  {Petitjean}, {Pieri}, {Pinsonneault}, {Porto de Mello}, {Prada}, {Prakash},
  {Price-Whelan}, {Protopapas}, {Raddick}, {Rahman}, {Reid}, {Rich}, {Rix},
  {Robin}, {Rockosi}, {Rodrigues}, {Rodr{\'\i}guez-Torres}, {Roe}, {Ross},
  {Ross}, {Rossi}, {Ruan}, {Rubi{\~n}o-Mart{\'\i}n}, {Rykoff},
  {Salazar-Albornoz}, {Salvato}, {Samushia}, {S{\'a}nchez}, {Santiago},
  {Sayres}, {Schiavon}, {Schlegel}, {Schmidt}, {Schneider}, {Schultheis},
  {Schwope}, {Sc{\'o}ccola}, {Scott}, {Sellgren}, {Seo}, {Serenelli}, {Shane},
  {Shen}, {Shetrone}, {Shu}, {Silva Aguirre}, {Sivarani}, {Skrutskie},
  {Slosar}, {Smith}, {Sobreira}, {Souto}, {Stassun}, {Steinmetz}, {Stello},
  {Strauss}, {Streblyanska}, {Suzuki}, {Swanson}, {Tan}, {Tayar}, {Terrien},
  {Thakar}, {Thomas}, {Thomas}, {Thompson}, {Tinker}, {Tojeiro}, {Troup},
  {Vargas-Maga{\~n}a}, {Vazquez}, {Verde}, {Viel}, {Vogt}, {Wake}, {Wang},
  {Weaver}, {Weinberg}, {Weiner}, {White}, {Wilson}, {Wisniewski},
  {Wood-Vasey}, {Ye`che}, {York}, {Zakamska}, {Zamora}, {Zasowski}, {Zehavi},
  {Zhao}, {Zheng}, {Zhou}, {Zhou}, {Zou}, \& {Zhu}}]{dr12}
{Alam}, S., {Albareti}, F.~D., {Allende Prieto}, C., {et~al.} 2015, \apjs, 219,
  12, \dodoi{10.1088/0067-0049/219/1/12}

\bibitem[{{Arrigoni Battaia} {et~al.}(2016){Arrigoni Battaia}, {Hennawi},
  {Cantalupo}, \& {Prochaska}}]{arrigoni-battaia2016}
{Arrigoni Battaia}, F., {Hennawi}, J.~F., {Cantalupo}, S., \& {Prochaska},
  J.~X. 2016, \apj, 829, 3, \dodoi{10.3847/0004-637X/829/1/3}

\bibitem[{{Arrigoni Battaia} {et~al.}(2019){Arrigoni Battaia}, {Hennawi},
  {Prochaska}, {O{\~n}orbe}, {Farina}, {Cantalupo}, \&
  {Lusso}}]{arrigoni-battaia2019}
{Arrigoni Battaia}, F., {Hennawi}, J.~F., {Prochaska}, J.~X., {et~al.} 2019,
  \mnras, 482, 3162, \dodoi{10.1093/mnras/sty2827}

\bibitem[{{Atek} {et~al.}(2014){Atek}, {Kunth}, {Schaerer}, {Mas-Hesse},
  {Hayes}, {{\"O}stlin}, \& {Kneib}}]{atek2014}
{Atek}, H., {Kunth}, D., {Schaerer}, D., {et~al.} 2014, \aap, 561, A89,
  \dodoi{10.1051/0004-6361/201321519}

\bibitem[{{Balashev} {et~al.}(2017){Balashev}, {Noterdaeme}, {Rahmani},
  {Klimenko}, {Ledoux}, {Petitjean}, {Srianand}, {Ivanchik}, \&
  {Varshalovich}}]{balashev2017}
{Balashev}, S.~A., {Noterdaeme}, P., {Rahmani}, H., {et~al.} 2017, \mnras, 470,
  2890, \dodoi{10.1093/mnras/stx1339}

\bibitem[{{Begum} {et~al.}(2008){Begum}, {Chengalur}, {Karachentsev},
  {Sharina}, \& {Kaisin}}]{begum2008}
{Begum}, A., {Chengalur}, J.~N., {Karachentsev}, I.~D., {Sharina}, M.~E., \&
  {Kaisin}, S.~S. 2008, \mnras, 386, 1667,
  \dodoi{10.1111/j.1365-2966.2008.13150.x}

\bibitem[{{Bera} {et~al.}(2019){Bera}, {Kanekar}, {Chengalur}, \&
  {Bagla}}]{bera2019}
{Bera}, A., {Kanekar}, N., {Chengalur}, J.~N., \& {Bagla}, J.~S. 2019, \apjl,
  882, L7, \dodoi{10.3847/2041-8213/ab3656}

\bibitem[{{Berg} {et~al.}(2015){Berg}, {Neeleman}, {Prochaska}, {Ellison}, \&
  {Wolfe}}]{berg2015}
{Berg}, T. A.~M., {Neeleman}, M., {Prochaska}, J.~X., {Ellison}, S.~L., \&
  {Wolfe}, A.~M. 2015, \pasp, 127, 167, \dodoi{10.1086/680210}

\bibitem[{{Blanc} {et~al.}(2011){Blanc}, {Adams}, {Gebhardt}, {Hill}, {Drory},
  {Hao}, {Bender}, {Ciardullo}, {Finkelstein}, {Fry}, {Gawiser}, {Gronwall},
  {Hopp}, {Jeong}, {Kelzenberg}, {Komatsu}, {MacQueen}, {Murphy}, {Roth},
  {Schneider}, \& {Tufts}}]{blanc2011}
{Blanc}, G.~A., {Adams}, J.~J., {Gebhardt}, K., {et~al.} 2011, \apj, 736, 31,
  \dodoi{10.1088/0004-637X/736/1/31}

\bibitem[{{Bolatto} {et~al.}(2013){Bolatto}, {Wolfire}, \&
  {Leroy}}]{bolatto2013}
{Bolatto}, A.~D., {Wolfire}, M., \& {Leroy}, A.~K. 2013, \araa, 51, 207,
  \dodoi{10.1146/annurev-astro-082812-140944}

\bibitem[{{Broeils} \& {Rhee}(1997)}]{broeils97}
{Broeils}, A.~H., \& {Rhee}, M.~H. 1997, \aap, 324, 877

\bibitem[{{Cai} {et~al.}(2019){Cai}, {Cantalupo}, {Prochaska}, {Arrigoni
  Battaia}, {Burchett}, {Li}, {Chisholm}, {Bundy}, \& {Hennawi}}]{zheng2019b}
{Cai}, Z., {Cantalupo}, S., {Prochaska}, J.~X., {et~al.} 2019, \apjs, 245, 23,
  \dodoi{10.3847/1538-4365/ab4796}

\bibitem[{{Cassata} {et~al.}(2015){Cassata}, {Tasca}, {Le F{\`e}vre}, {Lemaux},
  {Garilli}, {Le Brun}, {Maccagni}, {Pentericci}, {Thomas}, {Vanzella},
  {Zamorani}, {Zucca}, {Amorin}, {Bardelli}, {Capak}, {Cassar{\`a}},
  {Castellano}, {Cimatti}, {Cuby}, {Cucciati}, {de la Torre}, {Durkalec},
  {Fontana}, {Giavalisco}, {Grazian}, {Hathi}, {Ilbert}, {Moreau}, {Paltani},
  {Ribeiro}, {Salvato}, {Schaerer}, {Scodeggio}, {Sommariva}, {Talia},
  {Taniguchi}, {Tresse}, {Vergani}, {Wang}, {Charlot}, {Contini}, {Fotopoulou},
  {Koekemoer}, {L{\'o}pez-Sanjuan}, {Mellier}, \& {Scoville}}]{cassata2015}
{Cassata}, P., {Tasca}, L.~A.~M., {Le F{\`e}vre}, O., {et~al.} 2015, \aap, 573,
  A24, \dodoi{10.1051/0004-6361/201423824}

\bibitem[{{Catinella} {et~al.}(2018){Catinella}, {Saintonge}, {Janowiecki},
  {Cortese}, {Dav{\'e}}, {Lemonias}, {Cooper}, {Schiminovich}, {Hummels},
  {Fabello}, {Ger{\'e}b}, {Kilborn}, \& {Wang}}]{catinella2018}
{Catinella}, B., {Saintonge}, A., {Janowiecki}, S., {et~al.} 2018, \mnras, 476,
  875, \dodoi{10.1093/mnras/sty089}

\bibitem[{{Chabrier}(2003)}]{chabrier2003}
{Chabrier}, G. 2003, \pasp, 115, 763, \dodoi{10.1086/376392}

\bibitem[{{Charlot} \& {Fall}(1993)}]{charlot1993}
{Charlot}, S., \& {Fall}, S.~M. 1993, \apj, 415, 580, \dodoi{10.1086/173187}

\bibitem[{{Chowdhury} {et~al.}(2022{\natexlab{a}}){Chowdhury}, {Kanekar}, \&
  {Chengalur}}]{chowdhury2022}
{Chowdhury}, A., {Kanekar}, N., \& {Chengalur}, J.~N. 2022{\natexlab{a}}, \apj,
  937, 103, \dodoi{10.3847/1538-4357/ac7d52}

\bibitem[{{Chowdhury} {et~al.}(2022{\natexlab{b}}){Chowdhury}, {Kanekar}, \&
  {Chengalur}}]{chowdhury2022b}
---. 2022{\natexlab{b}}, \apjl, 935, L5, \dodoi{10.3847/2041-8213/ac8150}

\bibitem[{{Chowdhury} {et~al.}(2020){Chowdhury}, {Kanekar}, {Chengalur},
  {Sethi}, \& {Dwarakanath}}]{chowdhury2020}
{Chowdhury}, A., {Kanekar}, N., {Chengalur}, J.~N., {Sethi}, S., \&
  {Dwarakanath}, K.~S. 2020, \nat, 586, 369, \dodoi{10.1038/s41586-020-2794-7}

\bibitem[{{Chowdhury} {et~al.}(2021){Chowdhury}, {Kanekar}, {Das},
  {Dwarakanath}, \& {Sethi}}]{chowdhury2021}
{Chowdhury}, A., {Kanekar}, N., {Das}, B., {Dwarakanath}, K.~S., \& {Sethi}, S.
  2021, \apjl, 913, L24, \dodoi{10.3847/2041-8213/abfcc7}

\bibitem[{{Christensen} {et~al.}(2014){Christensen}, {M{\o}ller}, {Fynbo}, \&
  {Zafar}}]{christensen2014}
{Christensen}, L., {M{\o}ller}, P., {Fynbo}, J.~P.~U., \& {Zafar}, T. 2014,
  \mnras, 445, 225, \dodoi{10.1093/mnras/stu1726}

\bibitem[{{Christensen} {et~al.}(2007){Christensen}, {Wisotzki}, {Roth},
  {S{\'a}nchez}, {Kelz}, \& {Jahnke}}]{christensen2007}
{Christensen}, L., {Wisotzki}, L., {Roth}, M.~M., {et~al.} 2007, \aap, 468,
  587, \dodoi{10.1051/0004-6361:20066410}

\bibitem[{{Chung} {et~al.}(2009){Chung}, {van Gorkom}, {Kenney}, {Crowl}, \&
  {Vollmer}}]{chung2009}
{Chung}, A., {van Gorkom}, J.~H., {Kenney}, J. D.~P., {Crowl}, H., \&
  {Vollmer}, B. 2009, \aj, 138, 1741, \dodoi{10.1088/0004-6256/138/6/1741}

\bibitem[{{Ciardullo} {et~al.}(2014){Ciardullo}, {Zeimann}, {Gronwall},
  {Gebhardt}, {Schneider}, {Hagen}, {Malz}, {Blanc}, {Hill}, {Drory}, \&
  {Gawiser}}]{ciardullo2014}
{Ciardullo}, R., {Zeimann}, G.~R., {Gronwall}, C., {et~al.} 2014, \apj, 796,
  64, \dodoi{10.1088/0004-637X/796/1/64}

\bibitem[{{Cortese} {et~al.}(2021){Cortese}, {Catinella}, \&
  {Smith}}]{cortese2021}
{Cortese}, L., {Catinella}, B., \& {Smith}, R. 2021, \pasa, 38, e035,
  \dodoi{10.1017/pasa.2021.18}

\bibitem[{{Crighton} {et~al.}(2015){Crighton}, {Murphy}, {Prochaska},
  {Worseck}, {Rafelski}, {Becker}, {Ellison}, {Fumagalli}, {Lopez}, {Meiksin},
  \& {O'Meara}}]{crighton2015}
{Crighton}, N. H.~M., {Murphy}, M.~T., {Prochaska}, J.~X., {et~al.} 2015,
  \mnras, 452, 217, \dodoi{10.1093/mnras/stv1182}

\bibitem[{{Dijkstra} \& {Kramer}(2012)}]{dijkstra2012}
{Dijkstra}, M., \& {Kramer}, R. 2012, \mnras, 424, 1672,
  \dodoi{10.1111/j.1365-2966.2012.21131.x}

\bibitem[{{Duval} {et~al.}(2014){Duval}, {Schaerer}, {{\"O}stlin}, \&
  {Laursen}}]{duval2014}
{Duval}, F., {Schaerer}, D., {{\"O}stlin}, G., \& {Laursen}, P. 2014, \aap,
  562, A52, \dodoi{10.1051/0004-6361/201220455}

\bibitem[{{Erb} {et~al.}(2006){Erb}, {Shapley}, {Pettini}, {Steidel}, {Reddy},
  \& {Adelberger}}]{erb2006}
{Erb}, D.~K., {Shapley}, A.~E., {Pettini}, M., {et~al.} 2006, \apj, 644, 813,
  \dodoi{10.1086/503623}

\bibitem[{{Farina} {et~al.}(2017){Farina}, {Venemans}, {Decarli}, {Hennawi},
  {Walter}, {Ba{\~n}ados}, {Mazzucchelli}, {Cantalupo}, {Arrigoni-Battaia}, \&
  {McGreer}}]{farina2017}
{Farina}, E.~P., {Venemans}, B.~P., {Decarli}, R., {et~al.} 2017, \apj, 848,
  78, \dodoi{10.3847/1538-4357/aa8df4}

\bibitem[{{Farina} {et~al.}(2019){Farina}, {Arrigoni-Battaia}, {Costa},
  {Walter}, {Hennawi}, {Drake}, {Decarli}, {Gutcke}, {Mazzucchelli},
  {Neeleman}, {Georgiev}, {Eilers}, {Davies}, {Ba{\~n}ados}, {Fan}, {Onoue},
  {Schindler}, {Venemans}, {Wang}, {Yang}, {Rabien}, \& {Busoni}}]{farina2019}
{Farina}, E.~P., {Arrigoni-Battaia}, F., {Costa}, T., {et~al.} 2019, \apj, 887,
  196, \dodoi{10.3847/1538-4357/ab5847}

\bibitem[{{Fern{\'a}ndez} {et~al.}(2016){Fern{\'a}ndez}, {Gim}, {van Gorkom},
  {Yun}, {Momjian}, {Popping}, {Chomiuk}, {Hess}, {Hunt}, {Kreckel}, {Lucero},
  {Maddox}, {Oosterloo}, {Pisano}, {Verheijen}, {Hales}, {Chung}, {Dodson},
  {Golap}, {Gross}, {Henning}, {Hibbard}, {Jaff{\'e}}, {Donovan Meyer},
  {Meyer}, {Sanchez-Barrantes}, {Schiminovich}, {Wicenec}, {Wilcots},
  {Bershady}, {Scoville}, {Strader}, {Tremou}, {Salinas}, \&
  {Ch{\'a}vez}}]{fernandez2016}
{Fern{\'a}ndez}, X., {Gim}, H.~B., {van Gorkom}, J.~H., {et~al.} 2016, \apjl,
  824, L1, \dodoi{10.3847/2041-8205/824/1/L1}

\bibitem[{{Finlator} {et~al.}(2007){Finlator}, {Dav{\'e}}, \&
  {Oppenheimer}}]{finlator2007}
{Finlator}, K., {Dav{\'e}}, R., \& {Oppenheimer}, B.~D. 2007, \mnras, 376,
  1861, \dodoi{10.1111/j.1365-2966.2007.11578.x}

\bibitem[{{Fitzpatrick}(1999)}]{fitzpatrick1999}
{Fitzpatrick}, E.~L. 1999, \pasp, 111, 63, \dodoi{10.1086/316293}

\bibitem[{{Fumagalli} {et~al.}(2015){Fumagalli}, {O'Meara}, {Prochaska},
  {Rafelski}, \& {Kanekar}}]{fumagalli2015}
{Fumagalli}, M., {O'Meara}, J.~M., {Prochaska}, J.~X., {Rafelski}, M., \&
  {Kanekar}, N. 2015, \mnras, 446, 3178, \dodoi{10.1093/mnras/stu2325}

\bibitem[{{Fumagalli} {et~al.}(2017){Fumagalli}, {Mackenzie}, {Trayford},
  {Theuns}, {Cantalupo}, {Christensen}, {Fynbo}, {M{\o}ller}, {O'Meara},
  {Prochaska}, {Rafelski}, \& {Shanks}}]{fumagalli2017}
{Fumagalli}, M., {Mackenzie}, R., {Trayford}, J., {et~al.} 2017, \mnras, 471,
  3686, \dodoi{10.1093/mnras/stx1896}

\bibitem[{{Fynbo} {et~al.}(2003){Fynbo}, {Ledoux}, {M{\o}ller}, {Thomsen}, \&
  {Burud}}]{fynbo2003}
{Fynbo}, J.~P.~U., {Ledoux}, C., {M{\o}ller}, P., {Thomsen}, B., \& {Burud}, I.
  2003, \aap, 407, 147, \dodoi{10.1051/0004-6361:20030840}

\bibitem[{{Fynbo} {et~al.}(2008){Fynbo}, {Prochaska}, {Sommer-Larsen},
  {Dessauges-Zavadsky}, \& {M{\o}ller}}]{fynbo2008}
{Fynbo}, J. P.~U., {Prochaska}, J.~X., {Sommer-Larsen}, J.,
  {Dessauges-Zavadsky}, M., \& {M{\o}ller}, P. 2008, \apj, 683, 321,
  \dodoi{10.1086/589555}

\bibitem[{{Fynbo} {et~al.}(2010){Fynbo}, {Laursen}, {Ledoux}, {M{\o}ller},
  {Durgapal}, {Goldoni}, {Gullberg}, {Kaper}, {Maund}, {Noterdaeme},
  {{\"O}stlin}, {Strandet}, {Toft}, {Vreeswijk}, \& {Zafar}}]{fynbo2010}
{Fynbo}, J.~P.~U., {Laursen}, P., {Ledoux}, C., {et~al.} 2010, \mnras, 408,
  2128, \dodoi{10.1111/j.1365-2966.2010.17294.x}

\bibitem[{{Fynbo} {et~al.}(2011){Fynbo}, {Ledoux}, {Noterdaeme}, {Christensen},
  {M{\o}ller}, {Durgapal}, {Goldoni}, {Kaper}, {Krogager}, {Laursen}, {Maund},
  {Milvang-Jensen}, {Okoshi}, {Rasmussen}, {Thorsen}, {Toft}, \&
  {Zafar}}]{fynbo2011}
{Fynbo}, J.~P.~U., {Ledoux}, C., {Noterdaeme}, P., {et~al.} 2011, \mnras, 413,
  2481, \dodoi{10.1111/j.1365-2966.2011.18318.x}

\bibitem[{{Fynbo} {et~al.}(2013){Fynbo}, {Geier}, {Christensen}, {Gallazzi},
  {Krogager}, {Kr{\"u}hler}, {Ledoux}, {Maund}, {M{\o}ller}, {Noterdaeme},
  {Rivera-Thorsen}, \& {Vestergaard}}]{fynbo2013}
{Fynbo}, J.~P.~U., {Geier}, S.~J., {Christensen}, L., {et~al.} 2013, \mnras,
  436, 361, \dodoi{10.1093/mnras/stt1579}

\bibitem[{{Fynbo} {et~al.}(2018){Fynbo}, {Heintz}, {Neeleman}, {Christensen},
  {Dessauges-Zavadsky}, {Kanekar}, {M{\o}ller}, {Prochaska}, {Rhodin}, \&
  {Zwaan}}]{fynbo2018}
{Fynbo}, J.~P.~U., {Heintz}, K.~E., {Neeleman}, M., {et~al.} 2018, \mnras, 479,
  2126, \dodoi{10.1093/mnras/sty1520}

\bibitem[{{Fynbo} {et~al.}(2023){Fynbo}, {Christensen}, {Geier}, {Heintz},
  {Krogager}, {Ledoux}, {Milvang-Jensen}, {M{\o}ller}, {Vejlgaard}, {Viuho}, \&
  {{\"O}stlin}}]{fynbo2023}
{Fynbo}, J.~P.~U., {Christensen}, L., {Geier}, S.~J., {et~al.} 2023, \aap, 679,
  A30, \dodoi{10.1051/0004-6361/202347403}

\bibitem[{{Gaia Collaboration} {et~al.}(2023){Gaia Collaboration}, {Vallenari},
  {Brown}, {Prusti}, {de Bruijne}, {Arenou}, {Babusiaux}, {Biermann},
  {Creevey}, {Ducourant}, {Evans}, {Eyer}, {Guerra}, {Hutton}, {Jordi},
  {Klioner}, {Lammers}, {Lindegren}, {Luri}, {Mignard}, {Panem}, {Pourbaix},
  {Randich}, {Sartoretti}, {Soubiran}, {Tanga}, {Walton}, {Bailer-Jones},
  {Bastian}, {Drimmel}, {Jansen}, {Katz}, {Lattanzi}, {van Leeuwen}, {Bakker},
  {Cacciari}, {Casta{\~n}eda}, {De Angeli}, {Fabricius}, {Fouesneau},
  {Fr{\'e}mat}, {Galluccio}, {Guerrier}, {Heiter}, {Masana}, {Messineo},
  {Mowlavi}, {Nicolas}, {Nienartowicz}, {Pailler}, {Panuzzo}, {Riclet}, {Roux},
  {Seabroke}, {Sordo}, {Th{\'e}venin}, {Gracia-Abril}, {Portell}, {Teyssier},
  {Altmann}, {Andrae}, {Audard}, {Bellas-Velidis}, {Benson}, {Berthier},
  {Blomme}, {Burgess}, {Busonero}, {Busso}, {C{\'a}novas}, {Carry}, {Cellino},
  {Cheek}, {Clementini}, {Damerdji}, {Davidson}, {de Teodoro}, {Nu{\~n}ez
  Campos}, {Delchambre}, {Dell'Oro}, {Esquej}, {Fern{\'a}ndez-Hern{\'a}ndez},
  {Fraile}, {Garabato}, {Garc{\'\i}a-Lario}, {Gosset}, {Haigron}, {Halbwachs},
  {Hambly}, {Harrison}, {Hern{\'a}ndez}, {Hestroffer}, {Hodgkin}, {Holl},
  {Jan{\ss}en}, {Jevardat de Fombelle}, {Jordan}, {Krone-Martins}, {Lanzafame},
  {L{\"o}ffler}, {Marchal}, {Marrese}, {Moitinho}, {Muinonen}, {Osborne},
  {Pancino}, {Pauwels}, {Recio-Blanco}, {Reyl{\'e}}, {Riello}, {Rimoldini},
  {Roegiers}, {Rybizki}, {Sarro}, {Siopis}, {Smith}, {Sozzetti}, {Utrilla},
  {van Leeuwen}, {Abbas}, {{\'A}brah{\'a}m}, {Abreu Aramburu}, {Aerts},
  {Aguado}, {Ajaj}, {Aldea-Montero}, {Altavilla}, {{\'A}lvarez}, {Alves},
  {Anders}, {Anderson}, {Anglada Varela}, {Antoja}, {Baines}, {Baker},
  {Balaguer-N{\'u}{\~n}ez}, {Balbinot}, {Balog}, {Barache}, {Barbato},
  {Barros}, {Barstow}, {Bartolom{\'e}}, {Bassilana}, {Bauchet}, {Becciani},
  {Bellazzini}, {Berihuete}, {Bernet}, {Bertone}, {Bianchi}, {Binnenfeld},
  {Blanco-Cuaresma}, {Blazere}, {Boch}, {Bombrun}, {Bossini}, {Bouquillon},
  {Bragaglia}, {Bramante}, {Breedt}, {Bressan}, {Brouillet}, {Brugaletta},
  {Bucciarelli}, {Burlacu}, {Butkevich}, {Buzzi}, {Caffau}, {Cancelliere},
  {Cantat-Gaudin}, {Carballo}, {Carlucci}, {Carnerero}, {Carrasco},
  {Casamiquela}, {Castellani}, {Castro-Ginard}, {Chaoul}, {Charlot}, {Chemin},
  {Chiaramida}, {Chiavassa}, {Chornay}, {Comoretto}, {Contursi}, {Cooper},
  {Cornez}, {Cowell}, {Crifo}, {Cropper}, {Crosta}, {Crowley}, {Dafonte},
  {Dapergolas}, {David}, {David}, {de Laverny}, {De Luise}, {De March}, {De
  Ridder}, {de Souza}, {de Torres}, {del Peloso}, {del Pozo}, {Delbo},
  {Delgado}, {Delisle}, {Demouchy}, {Dharmawardena}, {Di Matteo}, {Diakite},
  {Diener}, {Distefano}, {Dolding}, {Edvardsson}, {Enke}, {Fabre}, {Fabrizio},
  {Faigler}, {Fedorets}, {Fernique}, {Fienga}, {Figueras}, {Fournier},
  {Fouron}, {Fragkoudi}, {Gai}, {Garcia-Gutierrez}, {Garcia-Reinaldos},
  {Garc{\'\i}a-Torres}, {Garofalo}, {Gavel}, {Gavras}, {Gerlach}, {Geyer},
  {Giacobbe}, {Gilmore}, {Girona}, {Giuffrida}, {Gomel}, {Gomez},
  {Gonz{\'a}lez-N{\'u}{\~n}ez}, {Gonz{\'a}lez-Santamar{\'\i}a},
  {Gonz{\'a}lez-Vidal}, {Granvik}, {Guillout}, {Guiraud},
  {Guti{\'e}rrez-S{\'a}nchez}, {Guy}, {Hatzidimitriou}, {Hauser}, {Haywood},
  {Helmer}, {Helmi}, {Sarmiento}, {Hidalgo}, {Hilger}, {H{\l}adczuk}, {Hobbs},
  {Holland}, {Huckle}, {Jardine}, {Jasniewicz}, {Jean-Antoine Piccolo},
  {Jim{\'e}nez-Arranz}, {Jorissen}, {Juaristi Campillo}, {Julbe}, {Karbevska},
  {Kervella}, {Khanna}, {Kontizas}, {Kordopatis}, {Korn}, {K{\'o}sp{\'a}l},
  {Kostrzewa-Rutkowska}, {Kruszy{\'n}ska}, {Kun}, {Laizeau}, {Lambert},
  {Lanza}, {Lasne}, {Le Campion}, {Lebreton}, {Lebzelter}, {Leccia}, {Leclerc},
  {Lecoeur-Taibi}, {Liao}, {Licata}, {Lindstr{\o}m}, {Lister}, {Livanou},
  {Lobel}, {Lorca}, {Loup}, {Madrero Pardo}, {Magdaleno Romeo}, {Managau},
  {Mann}, {Manteiga}, {Marchant}, {Marconi}, {Marcos}, {Marcos Santos},
  {Mar{\'\i}n Pina}, {Marinoni}, {Marocco}, {Marshall}, {Martin Polo},
  {Mart{\'\i}n-Fleitas}, {Marton}, {Mary}, {Masip}, {Massari},
  {Mastrobuono-Battisti}, {Mazeh}, {McMillan}, {Messina}, {Michalik}, {Millar},
  {Mints}, {Molina}, {Molinaro}, {Moln{\'a}r}, {Monari}, {Mongui{\'o}},
  {Montegriffo}, {Montero}, {Mor}, {Mora}, {Morbidelli}, {Morel}, {Morris},
  {Muraveva}, {Murphy}, {Musella}, {Nagy}, {Noval}, {Oca{\~n}a}, {Ogden},
  {Ordenovic}, {Osinde}, {Pagani}, {Pagano}, {Palaversa}, {Palicio},
  {Pallas-Quintela}, {Panahi}, {Payne-Wardenaar}, {Pe{\~n}alosa Esteller},
  {Penttil{\"a}}, {Pichon}, {Piersimoni}, {Pineau}, {Plachy}, {Plum}, {Poggio},
  {Pr{\v{s}}a}, {Pulone}, {Racero}, {Ragaini}, {Rainer}, {Raiteri}, {Rambaux},
  {Ramos}, {Ramos-Lerate}, {Re Fiorentin}, {Regibo}, {Richards}, {Rios Diaz},
  {Ripepi}, {Riva}, {Rix}, {Rixon}, {Robichon}, {Robin}, {Robin}, {Roelens},
  {Rogues}, {Rohrbasser}, {Romero-G{\'o}mez}, {Rowell}, {Royer}, {Ruz Mieres},
  {Rybicki}, {Sadowski}, {S{\'a}ez N{\'u}{\~n}ez}, {Sagrist{\`a} Sell{\'e}s},
  {Sahlmann}, {Salguero}, {Samaras}, {Sanchez Gimenez}, {Sanna},
  {Santove{\~n}a}, {Sarasso}, {Schultheis}, {Sciacca}, {Segol}, {Segovia},
  {S{\'e}gransan}, {Semeux}, {Shahaf}, {Siddiqui}, {Siebert}, {Siltala},
  {Silvelo}, {Slezak}, {Slezak}, {Smart}, {Snaith}, {Solano}, {Solitro},
  {Souami}, {Souchay}, {Spagna}, {Spina}, {Spoto}, {Steele},
  {Steidelm{\"u}ller}, {Stephenson}, {S{\"u}veges}, {Surdej}, {Szabados},
  {Szegedi-Elek}, {Taris}, {Taylor}, {Teixeira}, {Tolomei}, {Tonello}, {Torra},
  {Torra}, {Torralba Elipe}, {Trabucchi}, {Tsounis}, {Turon}, {Ulla}, {Unger},
  {Vaillant}, {van Dillen}, {van Reeven}, {Vanel}, {Vecchiato}, {Viala},
  {Vicente}, {Voutsinas}, {Weiler}, {Wevers}, {Wyrzykowski}, {Yoldas}, {Yvard},
  {Zhao}, {Zorec}, {Zucker}, \& {Zwitter}}]{gaiadr3}
{Gaia Collaboration}, {Vallenari}, A., {Brown}, A.~G.~A., {et~al.} 2023, \aap,
  674, A1, \dodoi{10.1051/0004-6361/202243940}

\bibitem[{{Gallazzi} {et~al.}(2021){Gallazzi}, {Pasquali}, {Zibetti}, \&
  {Barbera}}]{gallazzi2020}
{Gallazzi}, A.~R., {Pasquali}, A., {Zibetti}, S., \& {Barbera}, F.~L. 2021,
  \mnras, 502, 4457, \dodoi{10.1093/mnras/stab265}

\bibitem[{{Giavalisco} {et~al.}(1994){Giavalisco}, {Macchetto}, \&
  {Sparks}}]{giavalisco1994}
{Giavalisco}, M., {Macchetto}, F.~D., \& {Sparks}, W.~B. 1994, \aap, 288, 103

\bibitem[{{Gronke} \& {Dijkstra}(2016)}]{gronke2016}
{Gronke}, M., \& {Dijkstra}, M. 2016, \apj, 826, 14,
  \dodoi{10.3847/0004-637X/826/1/14}

\bibitem[{{Gronke} {et~al.}(2016){Gronke}, {Dijkstra}, {McCourt}, \&
  {Oh}}]{gronke2016b}
{Gronke}, M., {Dijkstra}, M., {McCourt}, M., \& {Oh}, S.~P. 2016, \apjl, 833,
  L26, \dodoi{10.3847/2041-8213/833/2/L26}

\bibitem[{{Gronwall} {et~al.}(2007){Gronwall}, {Ciardullo}, {Hickey},
  {Gawiser}, {Feldmeier}, {van Dokkum}, {Urry}, {Herrera}, {Lehmer}, {Infante},
  {Orsi}, {Marchesini}, {Blanc}, {Francke}, {Lira}, \&
  {Treister}}]{gronwall2007}
{Gronwall}, C., {Ciardullo}, R., {Hickey}, T., {et~al.} 2007, \apj, 667, 79,
  \dodoi{10.1086/520324}

\bibitem[{{Grove} {et~al.}(2009){Grove}, {Fynbo}, {Ledoux}, {Limousin},
  {M{\o}ller}, {Nilsson}, \& {Thomsen}}]{grove2009}
{Grove}, L.~F., {Fynbo}, J.~P.~U., {Ledoux}, C., {et~al.} 2009, \aap, 497, 689,
  \dodoi{10.1051/0004-6361/200811429}

\bibitem[{{Guaita} {et~al.}(2010){Guaita}, {Gawiser}, {Padilla}, {Francke},
  {Bond}, {Gronwall}, {Ciardullo}, {Feldmeier}, {Sinawa}, {Blanc}, \&
  {Virani}}]{guaita2010}
{Guaita}, L., {Gawiser}, E., {Padilla}, N., {et~al.} 2010, \apj, 714, 255,
  \dodoi{10.1088/0004-637X/714/1/255}

\bibitem[{{Gunn} \& {Gott}(1972)}]{gunn-gott1972}
{Gunn}, J.~E., \& {Gott}, J.~Richard, I. 1972, \apj, 176, 1,
  \dodoi{10.1086/151605}

\bibitem[{{Hayes} {et~al.}(2011){Hayes}, {Schaerer}, {{\"O}stlin}, {Mas-Hesse},
  {Atek}, \& {Kunth}}]{hayes2011}
{Hayes}, M., {Schaerer}, D., {{\"O}stlin}, G., {et~al.} 2011, \apj, 730, 8,
  \dodoi{10.1088/0004-637X/730/1/8}

\bibitem[{{Haynes} {et~al.}(2018){Haynes}, {Giovanelli}, {Kent}, {Adams},
  {Balonek}, {Craig}, {Fertig}, {Finn}, {Giovanardi}, {Hallenbeck}, {Hess},
  {Hoffman}, {Huang}, {Jones}, {Koopmann}, {Kornreich}, {Leisman}, {Miller},
  {Moorman}, {O'Connor}, {O'Donoghue}, {Papastergis}, {Troischt}, {Stark}, \&
  {Xiao}}]{haynes2018}
{Haynes}, M.~P., {Giovanelli}, R., {Kent}, B.~R., {et~al.} 2018, \apj, 861, 49,
  \dodoi{10.3847/1538-4357/aac956}

\bibitem[{{Heald} {et~al.}(2011){Heald}, {J{\'o}zsa}, {Serra}, {Zschaechner},
  {Rand}, {Fraternali}, {Oosterloo}, {Walterbos}, {J{\"u}tte}, \&
  {Gentile}}]{heald2011}
{Heald}, G., {J{\'o}zsa}, G., {Serra}, P., {et~al.} 2011, \aap, 526, A118,
  \dodoi{10.1051/0004-6361/201015938}

\bibitem[{{Herenz} {et~al.}(2015){Herenz}, {Wisotzki}, {Roth}, \&
  {Anders}}]{herenz2015}
{Herenz}, E.~C., {Wisotzki}, L., {Roth}, M., \& {Anders}, F. 2015, \aap, 576,
  A115, \dodoi{10.1051/0004-6361/201425580}

\bibitem[{{Jiang} {et~al.}(2013){Jiang}, {Egami}, {Mechtley}, {Fan}, {Cohen},
  {Windhorst}, {Dav{\'e}}, {Finlator}, {Kashikawa}, {Ouchi}, \&
  {Shimasaku}}]{jiang2013a}
{Jiang}, L., {Egami}, E., {Mechtley}, M., {et~al.} 2013, \apj, 772, 99,
  \dodoi{10.1088/0004-637X/772/2/99}

\bibitem[{{Jones} {et~al.}(2018){Jones}, {Haynes}, {Giovanelli}, \&
  {Moorman}}]{jones2018}
{Jones}, M.~G., {Haynes}, M.~P., {Giovanelli}, R., \& {Moorman}, C. 2018,
  \mnras, 477, 2, \dodoi{10.1093/mnras/sty521}

\bibitem[{{Jorgenson} {et~al.}(2013){Jorgenson}, {Murphy}, \&
  {Thompson}}]{jorgenson2013}
{Jorgenson}, R.~A., {Murphy}, M.~T., \& {Thompson}, R. 2013, \mnras, 435, 482,
  \dodoi{10.1093/mnras/stt1309}

\bibitem[{{Jorgenson} \& {Wolfe}(2014)}]{jorgenson2014}
{Jorgenson}, R.~A., \& {Wolfe}, A.~M. 2014, \apj, 785, 16,
  \dodoi{10.1088/0004-637X/785/1/16}

\bibitem[{{Joshi} {et~al.}(2021){Joshi}, {Fumagalli}, {Srianand}, {Noterdaeme},
  {Petitjean}, {Rafelski}, {Mackenzie}, {Li}, {Cai}, {Martin}, {Zou}, {Wu},
  {Jiang}, \& {Ho}}]{joshi2021}
{Joshi}, R., {Fumagalli}, M., {Srianand}, R., {et~al.} 2021, \apj, 908, 129,
  \dodoi{10.3847/1538-4357/abd1d5}

\bibitem[{{Kanekar} {et~al.}(2020){Kanekar}, {Prochaska}, {Neeleman},
  {Christensen}, {M{\o}ller}, {Zwaan}, {Fynbo}, \&
  {Dessauges-Zavadsky}}]{kanekar2020}
{Kanekar}, N., {Prochaska}, J.~X., {Neeleman}, M., {et~al.} 2020, \apjl, 901,
  L5, \dodoi{10.3847/2041-8213/abb4e1}

\bibitem[{{Kanekar} {et~al.}(2014){Kanekar}, {Prochaska}, {Smette}, {Ellison},
  {Ryan-Weber}, {Momjian}, {Briggs}, {Lane}, {Chengalur}, {Delafosse}, {Grave},
  {Jacobsen}, \& {de Bruyn}}]{kanekar2014}
{Kanekar}, N., {Prochaska}, J.~X., {Smette}, A., {et~al.} 2014, \mnras, 438,
  2131, \dodoi{10.1093/mnras/stt2338}

\bibitem[{{Kaplan} {et~al.}(2010){Kaplan}, {Prochaska}, {Herbert-Fort},
  {Ellison}, \& {Dessauges-Zavadsky}}]{kaplan2010}
{Kaplan}, K.~F., {Prochaska}, J.~X., {Herbert-Fort}, S., {Ellison}, S.~L., \&
  {Dessauges-Zavadsky}, M. 2010, \pasp, 122, 619, \dodoi{10.1086/653500}

\bibitem[{{Kaur} {et~al.}(2022{\natexlab{a}}){Kaur}, {Kanekar}, {Rafelski},
  {Neeleman}, {Prochaska}, \& {Revalski}}]{kaur2022b}
{Kaur}, B., {Kanekar}, N., {Rafelski}, M., {et~al.} 2022{\natexlab{a}}, \apjl,
  933, L42, \dodoi{10.3847/2041-8213/ac7bdd}

\bibitem[{{Kaur} {et~al.}(2021){Kaur}, {Kanekar}, {Rafelski}, {Neeleman},
  {Revalski}, \& {Prochaska}}]{kaur2021}
---. 2021, \apj, 921, 68, \dodoi{10.3847/1538-4357/ac12d2}

\bibitem[{{Kaur} {et~al.}(2022{\natexlab{b}}){Kaur}, {Kanekar}, {Revalski},
  {Rafelski}, {Neeleman}, {Prochaska}, \& {Walter}}]{kaur2022a}
{Kaur}, B., {Kanekar}, N., {Revalski}, M., {et~al.} 2022{\natexlab{b}}, \apj,
  934, 87, \dodoi{10.3847/1538-4357/ac7b2c}

\bibitem[{{Kawata} \& {Mulchaey}(2008)}]{kawata-mulchaey2008}
{Kawata}, D., \& {Mulchaey}, J.~S. 2008, \apjl, 672, L103,
  \dodoi{10.1086/526544}

\bibitem[{{Kennicutt} \& {Evans}(2012)}]{kennicutt2012}
{Kennicutt}, R.~C., \& {Evans}, N.~J. 2012, \araa, 50, 531,
  \dodoi{10.1146/annurev-astro-081811-125610}

\bibitem[{{Kere{\v s}} {et~al.}(2005){Kere{\v s}}, {Katz}, {Weinberg}, \&
  {Dav{\'e}}}]{keres2005}
{Kere{\v s}}, D., {Katz}, N., {Weinberg}, D.~H., \& {Dav{\'e}}, R. 2005,
  \mnras, 363, 2, \dodoi{10.1111/j.1365-2966.2005.09451.x}

\bibitem[{{Klimenko} {et~al.}(2020){Klimenko}, {Petitjean}, \&
  {Ivanchik}}]{klimenko2020}
{Klimenko}, V.~V., {Petitjean}, P., \& {Ivanchik}, A.~V. 2020, \mnras, 493,
  5743, \dodoi{10.1093/mnras/staa614}

\bibitem[{{Krogager} {et~al.}(2016){Krogager}, {Fynbo}, {Noterdaeme}, {Zafar},
  {M{\o}ller}, {Ledoux}, {Kr{\"u}hler}, \& {Stockton}}]{krogager2016}
{Krogager}, J.~K., {Fynbo}, J.~P.~U., {Noterdaeme}, P., {et~al.} 2016, \mnras,
  455, 2698, \dodoi{10.1093/mnras/stv2346}

\bibitem[{{Krogager} {et~al.}(2017){Krogager}, {M{\o}ller}, {Fynbo}, \&
  {Noterdaeme}}]{krogager2017}
{Krogager}, J.~K., {M{\o}ller}, P., {Fynbo}, J.~P.~U., \& {Noterdaeme}, P.
  2017, \mnras, 469, 2959, \dodoi{10.1093/mnras/stx1011}

\bibitem[{{Krogager} {et~al.}(2013){Krogager}, {Fynbo}, {Ledoux},
  {Christensen}, {Gallazzi}, {Laursen}, {M{\o}ller}, {Noterdaeme},
  {P{\'e}roux}, {Pettini}, \& {Vestergaard}}]{krogager2013}
{Krogager}, J.-K., {Fynbo}, J. P.~U., {Ledoux}, C., {et~al.} 2013, \mnras, 433,
  3091, \dodoi{10.1093/mnras/stt955}

\bibitem[{{Kulkarni} {et~al.}(2006){Kulkarni}, {Woodgate}, {York}, {Thatte},
  {Meiring}, {Palunas}, \& {Wassell}}]{kulkarni2006}
{Kulkarni}, V.~P., {Woodgate}, B.~E., {York}, D.~G., {et~al.} 2006, \apj, 636,
  30, \dodoi{10.1086/497885}

\bibitem[{{Laursen} {et~al.}(2009){Laursen}, {Sommer-Larsen}, \&
  {Andersen}}]{laursen2009}
{Laursen}, P., {Sommer-Larsen}, J., \& {Andersen}, A.~C. 2009, \apj, 704, 1640,
  \dodoi{10.1088/0004-637X/704/2/1640}

\bibitem[{{Ledoux} {et~al.}(2006){Ledoux}, {Petitjean}, {Fynbo}, {M{\o}ller},
  \& {Srianand}}]{ledoux2006}
{Ledoux}, C., {Petitjean}, P., {Fynbo}, J.~P.~U., {M{\o}ller}, P., \&
  {Srianand}, R. 2006, \aap, 457, 71, \dodoi{10.1051/0004-6361:20054242}

\bibitem[{{Leroy} {et~al.}(2008){Leroy}, {Walter}, {Brinks}, {Bigiel}, {de
  Blok}, {Madore}, \& {Thornley}}]{leroy2008}
{Leroy}, A.~K., {Walter}, F., {Brinks}, E., {et~al.} 2008, \aj, 136, 2782,
  \dodoi{10.1088/0004-6256/136/6/2782}

\bibitem[{{Lofthouse} {et~al.}(2020){Lofthouse}, {Fumagalli}, {Fossati},
  {O'Meara}, {Murphy}, {Christensen}, {Prochaska}, {Cantalupo}, {Bielby},
  {Cooke}, {Lusso}, \& {Morris}}]{lofthouse2020}
{Lofthouse}, E.~K., {Fumagalli}, M., {Fossati}, M., {et~al.} 2020, \mnras, 491,
  2057, \dodoi{10.1093/mnras/stz3066}

\bibitem[{{Lofthouse} {et~al.}(2023){Lofthouse}, {Fumagalli}, {Fossati},
  {Dutta}, {Galbiati}, {Arrigoni Battaia}, {Cantalupo}, {Christensen}, {Cooke},
  {Longobardi}, {Murphy}, \& {Prochaska}}]{lofthouse2023}
---. 2023, \mnras, 518, 305, \dodoi{10.1093/mnras/stac3089}

\bibitem[{{Lowenthal} {et~al.}(1991){Lowenthal}, {Hogan}, {Green}, {Caulet},
  {Woodgate}, {Brown}, \& {Foltz}}]{lowenthal1991}
{Lowenthal}, J.~D., {Hogan}, C.~J., {Green}, R.~F., {et~al.} 1991, \apjl, 377,
  L73, \dodoi{10.1086/186120}

\bibitem[{{Mackenzie} {et~al.}(2019){Mackenzie}, {Fumagalli}, {Theuns},
  {Hatton}, {Garel}, {Cantalupo}, {Christensen}, {Fynbo}, {Kanekar},
  {M{\o}ller}, {O'Meara}, {Prochaska}, {Rafelski}, {Shanks}, \&
  {Trayford}}]{mackenzie2019}
{Mackenzie}, R., {Fumagalli}, M., {Theuns}, T., {et~al.} 2019, \mnras, 487,
  5070, \dodoi{10.1093/mnras/stz1501}

\bibitem[{{Matthee} {et~al.}(2016){Matthee}, {Sobral}, {Oteo}, {Best}, {Smail},
  {R{\"o}ttgering}, \& {Paulino-Afonso}}]{matthee2016}
{Matthee}, J., {Sobral}, D., {Oteo}, I., {et~al.} 2016, \mnras, 458, 449,
  \dodoi{10.1093/mnras/stw322}

\bibitem[{{Mawatari} {et~al.}(2012){Mawatari}, {Yamada}, {Nakamura},
  {Hayashino}, \& {Matsuda}}]{mawatari2012}
{Mawatari}, K., {Yamada}, T., {Nakamura}, Y., {Hayashino}, T., \& {Matsuda}, Y.
  2012, \apj, 759, 133, \dodoi{10.1088/0004-637X/759/2/133}

\bibitem[{{M{\'e}rida} {et~al.}(2023){M{\'e}rida}, {P{\'e}rez-Gonz{\'a}lez},
  {S{\'a}nchez-Bl{\'a}zquez}, {Garc{\'\i}a-Argum{\'a}nez}, {Annunziatella},
  {Costantin}, {Lumbreras-Calle}, {Alcalde-Pampliega}, {Barro},
  {Espino-Briones}, \& {Koekemoer}}]{merida2023}
{M{\'e}rida}, R.~M., {P{\'e}rez-Gonz{\'a}lez}, P.~G.,
  {S{\'a}nchez-Bl{\'a}zquez}, P., {et~al.} 2023, \apj, 950, 125,
  \dodoi{10.3847/1538-4357/acc7a3}

\bibitem[{{M{\o}ller} {et~al.}(2004){M{\o}ller}, {Fynbo}, \&
  {Fall}}]{moller2004}
{M{\o}ller}, P., {Fynbo}, J.~P.~U., \& {Fall}, S.~M. 2004, \aap, 422, L33,
  \dodoi{10.1051/0004-6361:20040194}

\bibitem[{{M{\o}ller} {et~al.}(2013){M{\o}ller}, {Fynbo}, {Ledoux}, \&
  {Nilsson}}]{moller2013}
{M{\o}ller}, P., {Fynbo}, J.~P.~U., {Ledoux}, C., \& {Nilsson}, K.~K. 2013,
  \mnras, 430, 2680, \dodoi{10.1093/mnras/stt067}

\bibitem[{{M{\o}ller} \& {Warren}(1993)}]{moller-warren1993}
{M{\o}ller}, P., \& {Warren}, S.~J. 1993, \aap, 270, 43

\bibitem[{{M{\o}ller} {et~al.}(2002){M{\o}ller}, {Warren}, {Fall}, {Fynbo}, \&
  {Jakobsen}}]{moller2002}
{M{\o}ller}, P., {Warren}, S.~J., {Fall}, S.~M., {Fynbo}, J.~U., \& {Jakobsen},
  P. 2002, \apj, 574, 51, \dodoi{10.1086/340934}

\bibitem[{{Morrissey} {et~al.}(2018){Morrissey}, {Matuszewski}, {Martin},
  {Neill}, {Epps}, {Fucik}, {Weber}, {Darvish}, {Adkins}, {Allen}, {Bartos},
  {Belicki}, {Cabak}, {Callahan}, {Cowley}, {Crabill}, {Deich}, {Delecroix},
  {Doppman}, {Hilyard}, {James}, {Kaye}, {Kokorowski}, {Kwok}, {Lanclos},
  {Milner}, {Moore}, {O'Sullivan}, {Parihar}, {Park}, {Phillips}, {Rizzi},
  {Rockosi}, {Rodriguez}, {Salaun}, {Seaman}, {Sheikh}, {Weiss}, \&
  {Zarzaca}}]{morrissey2018}
{Morrissey}, P., {Matuszewski}, M., {Martin}, D.~C., {et~al.} 2018, \apj, 864,
  93, \dodoi{10.3847/1538-4357/aad597}

\bibitem[{{Neeleman} {et~al.}(2018){Neeleman}, {Kanekar}, {Prochaska},
  {Christensen}, {Dessauges-Zavadsky}, {Fynbo}, {M{\o}ller}, \&
  {Zwaan}}]{neeleman2018}
{Neeleman}, M., {Kanekar}, N., {Prochaska}, J.~X., {et~al.} 2018, \apjl, 856,
  L12, \dodoi{10.3847/2041-8213/aab5b1}

\bibitem[{{Neeleman} {et~al.}(2017){Neeleman}, {Kanekar}, {Prochaska},
  {Rafelski}, {Carilli}, \& {Wolfe}}]{neeleman2017}
---. 2017, Science, 355, 1285, \dodoi{10.1126/science.aal1737}

\bibitem[{{Neeleman} {et~al.}(2019){Neeleman}, {Kanekar}, {Prochaska},
  {Rafelski}, \& {Carilli}}]{neeleman2019}
{Neeleman}, M., {Kanekar}, N., {Prochaska}, J.~X., {Rafelski}, M.~A., \&
  {Carilli}, C.~L. 2019, \apjl, 870, L19, \dodoi{10.3847/2041-8213/aaf871}

\bibitem[{{Neeleman} {et~al.}(2015){Neeleman}, {Prochaska}, \&
  {Wolfe}}]{neeleman2015}
{Neeleman}, M., {Prochaska}, J.~X., \& {Wolfe}, A.~M. 2015, \apj, 800, 7,
  \dodoi{10.1088/0004-637X/800/1/7}

\bibitem[{{Neeleman} {et~al.}(2013){Neeleman}, {Wolfe}, {Prochaska}, \&
  {Rafelski}}]{neeleman2013}
{Neeleman}, M., {Wolfe}, A.~M., {Prochaska}, J.~X., \& {Rafelski}, M. 2013,
  \apj, 769, 54, \dodoi{10.1088/0004-637X/769/1/54}

\bibitem[{{Nielsen} {et~al.}(2022){Nielsen}, {Kacprzak}, {Sameer}, {Murphy},
  {Nateghi}, {Charlton}, \& {Churchill}}]{nielsen2022}
{Nielsen}, N.~M., {Kacprzak}, G.~G., {Sameer}, {et~al.} 2022, \mnras, 514,
  6074, \dodoi{10.1093/mnras/stac1824}

\bibitem[{{Nilsson} {et~al.}(2009){Nilsson}, {M{\"o}ller-Nilsson}, {M{\o}ller},
  {Fynbo}, \& {Shapley}}]{nilsson2009}
{Nilsson}, K.~K., {M{\"o}ller-Nilsson}, O., {M{\o}ller}, P., {Fynbo}, J.~P.~U.,
  \& {Shapley}, A.~E. 2009, \mnras, 400, 232,
  \dodoi{10.1111/j.1365-2966.2009.15439.x}

\bibitem[{{Noterdaeme} {et~al.}(2008){Noterdaeme}, {Ledoux}, {Petitjean}, \&
  {Srianand}}]{noterdaeme2008}
{Noterdaeme}, P., {Ledoux}, C., {Petitjean}, P., \& {Srianand}, R. 2008, \aap,
  481, 327, \dodoi{10.1051/0004-6361:20078780}

\bibitem[{{Noterdaeme} {et~al.}(2012){Noterdaeme}, {Petitjean}, {Carithers},
  {P{\^a}ris}, {Font-Ribera}, {Bailey}, {Aubourg}, {Bizyaev}, {Ebelke},
  {Finley}, {Ge}, {Malanushenko}, {Malanushenko}, {Miralda-Escud{\'e}},
  {Myers}, {Oravetz}, {Pan}, {Pieri}, {Ross}, {Schneider}, {Simmons}, \&
  {York}}]{noterdaeme2012}
{Noterdaeme}, P., {Petitjean}, P., {Carithers}, W.~C., {et~al.} 2012, \aap,
  547, L1, \dodoi{10.1051/0004-6361/201220259}

\bibitem[{{Oke} \& {Gunn}(1983)}]{oke1983}
{Oke}, J.~B., \& {Gunn}, J.~E. 1983, \apj, 266, 713, \dodoi{10.1086/160817}

\bibitem[{{O'Sullivan} {et~al.}(2020){O'Sullivan}, {Martin}, {Matuszewski},
  {Hoadley}, {Hamden}, {Neill}, {Lin}, \& {Parihar}}]{osullivan2020}
{O'Sullivan}, D.~B., {Martin}, C., {Matuszewski}, M., {et~al.} 2020, \apj, 894,
  3, \dodoi{10.3847/1538-4357/ab838c}

\bibitem[{{Oyarz{\'u}n} {et~al.}(2017){Oyarz{\'u}n}, {Blanc}, {Gonz{\'a}lez},
  {Mateo}, \& {Bailey}}]{oyarzun2017}
{Oyarz{\'u}n}, G.~A., {Blanc}, G.~A., {Gonz{\'a}lez}, V., {Mateo}, M., \&
  {Bailey}, John~I., I. 2017, \apj, 843, 133, \dodoi{10.3847/1538-4357/aa7552}

\bibitem[{{Oyarz{\'u}n} {et~al.}(2023){Oyarz{\'u}n}, {Bundy}, {Westfall},
  {Lacerna}, {Yan}, {Brownstein}, {Drory}, \& {Lane}}]{oyarzun2023}
{Oyarz{\'u}n}, G.~A., {Bundy}, K., {Westfall}, K.~B., {et~al.} 2023, \apj, 947,
  13, \dodoi{10.3847/1538-4357/acbbca}

\bibitem[{{Oyarz{\'u}n} {et~al.}(2016){Oyarz{\'u}n}, {Blanc}, {Gonz{\'a}lez},
  {Mateo}, {Bailey}, {Finkelstein}, {Lira}, {Crane}, \&
  {Olszewski}}]{oyarzun2016}
{Oyarz{\'u}n}, G.~A., {Blanc}, G.~A., {Gonz{\'a}lez}, V., {et~al.} 2016, \apjl,
  821, L14, \dodoi{10.3847/2041-8205/821/1/L14}

\bibitem[{{Pasquali} {et~al.}(2010){Pasquali}, {Gallazzi}, {Fontanot}, {van den
  Bosch}, {De Lucia}, {Mo}, \& {Yang}}]{pasquali2010}
{Pasquali}, A., {Gallazzi}, A., {Fontanot}, F., {et~al.} 2010, \mnras, 407,
  937, \dodoi{10.1111/j.1365-2966.2010.17074.x}

\bibitem[{{P{\'e}roux} {et~al.}(2011){P{\'e}roux}, {Bouch{\'e}}, {Kulkarni},
  {York}, \& {Vladilo}}]{peroux2011}
{P{\'e}roux}, C., {Bouch{\'e}}, N., {Kulkarni}, V.~P., {York}, D.~G., \&
  {Vladilo}, G. 2011, \mnras, 410, 2237,
  \dodoi{10.1111/j.1365-2966.2010.17598.x}

\bibitem[{{P{\'e}roux} {et~al.}(2012){P{\'e}roux}, {Bouch{\'e}}, {Kulkarni},
  {York}, \& {Vladilo}}]{peroux2012}
---. 2012, \mnras, 419, 3060, \dodoi{10.1111/j.1365-2966.2011.19947.x}

\bibitem[{{Prochaska} {et~al.}(2005){Prochaska}, {Herbert-Fort}, \&
  {Wolfe}}]{prochaska2005}
{Prochaska}, J.~X., {Herbert-Fort}, S., \& {Wolfe}, A.~M. 2005, \apj, 635, 123,
  \dodoi{10.1086/497287}

\bibitem[{{Prochaska} {et~al.}(2019){Prochaska}, {Neeleman}, {Kanekar}, \&
  {Rafelski}}]{prochaska2019}
{Prochaska}, J.~X., {Neeleman}, M., {Kanekar}, N., \& {Rafelski}, M. 2019,
  \apjl, 886, L35, \dodoi{10.3847/2041-8213/ab55eb}

\bibitem[{{Prochaska} \& {Wolfe}(1997)}]{prochaska-wolfe1997}
{Prochaska}, J.~X., \& {Wolfe}, A.~M. 1997, \apj, 487, 73,
  \dodoi{10.1086/304591}

\bibitem[{{Prochaska} {et~al.}(2013){Prochaska}, {Hennawi}, {Lee}, {Cantalupo},
  {Bovy}, {Djorgovski}, {Ellison}, {Lau}, {Martin}, {Myers}, {Rubin}, \&
  {Simcoe}}]{prochaska2013}
{Prochaska}, J.~X., {Hennawi}, J.~F., {Lee}, K.-G., {et~al.} 2013, \apj, 776,
  136, \dodoi{10.1088/0004-637X/776/2/136}

\bibitem[{{Rafelski} {et~al.}(2014){Rafelski}, {Neeleman}, {Fumagalli},
  {Wolfe}, \& {Prochaska}}]{rafelski2014}
{Rafelski}, M., {Neeleman}, M., {Fumagalli}, M., {Wolfe}, A.~M., \&
  {Prochaska}, J.~X. 2014, \apjl, 782, L29, \dodoi{10.1088/2041-8205/782/2/L29}

\bibitem[{{Rafelski} {et~al.}(2012){Rafelski}, {Wolfe}, {Prochaska},
  {Neeleman}, \& {Mendez}}]{rafelski2012}
{Rafelski}, M., {Wolfe}, A.~M., {Prochaska}, J.~X., {Neeleman}, M., \&
  {Mendez}, A.~J. 2012, \apj, 755, 89, \dodoi{10.1088/0004-637X/755/2/89}

\bibitem[{{Rahmati} \& {Schaye}(2014)}]{rahmati-schaye2014}
{Rahmati}, A., \& {Schaye}, J. 2014, \mnras, 438, 529,
  \dodoi{10.1093/mnras/stt2235}

\bibitem[{{Rao} {et~al.}(2017){Rao}, {Turnshek}, {Sardane}, \&
  {Monier}}]{rao2017}
{Rao}, S.~M., {Turnshek}, D.~A., {Sardane}, G.~M., \& {Monier}, E.~M. 2017,
  \mnras, 471, 3428, \dodoi{10.1093/mnras/stx1787}

\bibitem[{{Reddy} {et~al.}(2006){Reddy}, {Steidel}, {Fadda}, {Yan}, {Pettini},
  {Shapley}, {Erb}, \& {Adelberger}}]{reddy2006}
{Reddy}, N.~A., {Steidel}, C.~C., {Fadda}, D., {et~al.} 2006, \apj, 644, 792,
  \dodoi{10.1086/503739}

\bibitem[{{Rhodin} {et~al.}(2019){Rhodin}, {Agertz}, {Christensen}, {Renaud},
  \& {Fynbo}}]{rhodin2019}
{Rhodin}, N.~H.~P., {Agertz}, O., {Christensen}, L., {Renaud}, F., \& {Fynbo},
  J.~P.~U. 2019, \mnras, 488, 3634, \dodoi{10.1093/mnras/stz1479}

\bibitem[{{Rivera-Thorsen} {et~al.}(2015){Rivera-Thorsen}, {Hayes},
  {{\"O}stlin}, {Duval}, {Orlitov{\'a}}, {Verhamme}, {Mas-Hesse}, {Schaerer},
  {Cannon}, {Ot{\'{\i}}-Floranes}, {Sandberg}, {Guaita}, {Adamo}, {Atek},
  {Herenz}, {Kunth}, {Laursen}, \& {Melinder}}]{rivera-thorsen2015}
{Rivera-Thorsen}, T.~E., {Hayes}, M., {{\"O}stlin}, G., {et~al.} 2015, \apj,
  805, 14, \dodoi{10.1088/0004-637X/805/1/14}

\bibitem[{{Roche} {et~al.}(2014){Roche}, {Humphrey}, \& {Binette}}]{roche2014}
{Roche}, N., {Humphrey}, A., \& {Binette}, L. 2014, \mnras, 443, 3795,
  \dodoi{10.1093/mnras/stu1430}

\bibitem[{{Schlafly} \& {Finkbeiner}(2011)}]{schlafly-finkbeiner2011}
{Schlafly}, E.~F., \& {Finkbeiner}, D.~P. 2011, \apj, 737, 103,
  \dodoi{10.1088/0004-637X/737/2/103}

\bibitem[{{Schlegel} {et~al.}(1998){Schlegel}, {Finkbeiner}, \&
  {Davis}}]{schlegel1998}
{Schlegel}, D.~J., {Finkbeiner}, D.~P., \& {Davis}, M. 1998, \apj, 500, 525,
  \dodoi{10.1086/305772}

\bibitem[{{Shapley} {et~al.}(2003){Shapley}, {Steidel}, {Pettini}, \&
  {Adelberger}}]{shapley2003}
{Shapley}, A.~E., {Steidel}, C.~C., {Pettini}, M., \& {Adelberger}, K.~L. 2003,
  \apj, 588, 65, \dodoi{10.1086/373922}

\bibitem[{{Sheinis} {et~al.}(2002){Sheinis}, {Bolte}, {Epps}, {Kibrick},
  {Miller}, {Radovan}, {Bigelow}, \& {Sutin}}]{Keck-ESI}
{Sheinis}, A.~I., {Bolte}, M., {Epps}, H.~W., {et~al.} 2002, \pasp, 114, 851,
  \dodoi{10.1086/341706}

\bibitem[{{Skelton} {et~al.}(2014){Skelton}, {Whitaker}, {Momcheva}, {Brammer},
  {van Dokkum}, {Labb{\'e}}, {Franx}, {van der Wel}, {Bezanson}, {Da Cunha},
  {Fumagalli}, {F{\"o}rster Schreiber}, {Kriek}, {Leja}, {Lundgren}, {Magee},
  {Marchesini}, {Maseda}, {Nelson}, {Oesch}, {Pacifici}, {Patel}, {Price},
  {Rix}, {Tal}, {Wake}, \& {Wuyts}}]{skelton2014}
{Skelton}, R.~E., {Whitaker}, K.~E., {Momcheva}, I.~G., {et~al.} 2014, \apjs,
  214, 24, \dodoi{10.1088/0067-0049/214/2/24}

\bibitem[{{Smith} {et~al.}(1989){Smith}, {Cohen}, {Burns}, {Moore}, \&
  {Uchida}}]{smith1989}
{Smith}, H.~E., {Cohen}, R.~D., {Burns}, J.~E., {Moore}, D.~J., \& {Uchida},
  B.~A. 1989, \apj, 347, 87, \dodoi{10.1086/168099}

\bibitem[{{Sobral} \& {Matthee}(2019)}]{sobral2019}
{Sobral}, D., \& {Matthee}, J. 2019, \aap, 623, A157,
  \dodoi{10.1051/0004-6361/201833075}

\bibitem[{{Stark} {et~al.}(2009){Stark}, {Ellis}, {Bunker}, {Bundy}, {Targett},
  {Benson}, \& {Lacy}}]{stark2009}
{Stark}, D.~P., {Ellis}, R.~S., {Bunker}, A., {et~al.} 2009, \apj, 697, 1493,
  \dodoi{10.1088/0004-637X/697/2/1493}

\bibitem[{{Steidel} {et~al.}(1999){Steidel}, {Adelberger}, {Giavalisco},
  {Dickinson}, \& {Pettini}}]{steidel1999}
{Steidel}, C.~C., {Adelberger}, K.~L., {Giavalisco}, M., {Dickinson}, M., \&
  {Pettini}, M. 1999, \apj, 519, 1, \dodoi{10.1086/307363}

\bibitem[{{Steidel} {et~al.}(2003){Steidel}, {Adelberger}, {Shapley},
  {Pettini}, {Dickinson}, \& {Giavalisco}}]{steidel2003}
{Steidel}, C.~C., {Adelberger}, K.~L., {Shapley}, A.~E., {et~al.} 2003, \apj,
  592, 728, \dodoi{10.1086/375772}

\bibitem[{{Steidel} {et~al.}(1996){Steidel}, {Giavalisco}, {Pettini},
  {Dickinson}, \& {Adelberger}}]{steidel1996}
{Steidel}, C.~C., {Giavalisco}, M., {Pettini}, M., {Dickinson}, M., \&
  {Adelberger}, K.~L. 1996, \apjl, 462, L17, \dodoi{10.1086/310029}

\bibitem[{{Stern} {et~al.}(2021){Stern}, {Sternberg}, {Faucher-Gigu{\`e}re},
  {Hafen}, {Fielding}, {Quataert}, {Wetzel}, {Angl{\'e}s-Alc{\'a}zar},
  {El-Badry}, {Kere{\v{s}}}, \& {Hopkins}}]{stern2021}
{Stern}, J., {Sternberg}, A., {Faucher-Gigu{\`e}re}, C.-A., {et~al.} 2021,
  \mnras, 507, 2869, \dodoi{10.1093/mnras/stab2240}

\bibitem[{{Tacconi} {et~al.}(2020){Tacconi}, {Genzel}, \&
  {Sternberg}}]{tacconi2020}
{Tacconi}, L.~J., {Genzel}, R., \& {Sternberg}, A. 2020, \araa, 58, 157,
  \dodoi{10.1146/annurev-astro-082812-141034}

\bibitem[{{Tremonti} {et~al.}(2004){Tremonti}, {Heckman}, {Kauffmann},
  {Brinchmann}, {Charlot}, {White}, {Seibert}, {Peng}, {Schlegel}, {Uomoto},
  {Fukugita}, \& {Brinkmann}}]{tremonti2004}
{Tremonti}, C.~A., {Heckman}, T.~M., {Kauffmann}, G., {et~al.} 2004, \apj, 613,
  898, \dodoi{10.1086/423264}

\bibitem[{{Treu} {et~al.}(2012){Treu}, {Trenti}, {Stiavelli}, {Auger}, \&
  {Bradley}}]{treu2012}
{Treu}, T., {Trenti}, M., {Stiavelli}, M., {Auger}, M.~W., \& {Bradley}, L.~D.
  2012, \apj, 747, 27, \dodoi{10.1088/0004-637X/747/1/27}

\bibitem[{{Trussler} {et~al.}(2021){Trussler}, {Maiolino}, {Maraston}, {Peng},
  {Thomas}, {Goddard}, \& {Lian}}]{trussler2021}
{Trussler}, J., {Maiolino}, R., {Maraston}, C., {et~al.} 2021, \mnras, 500,
  4469, \dodoi{10.1093/mnras/staa3545}

\bibitem[{{Tumlinson} {et~al.}(2017){Tumlinson}, {Peeples}, \&
  {Werk}}]{tumlinson2017}
{Tumlinson}, J., {Peeples}, M.~S., \& {Werk}, J.~K. 2017, \araa, 55, 389,
  \dodoi{10.1146/annurev-astro-091916-055240}

\bibitem[{{Verhamme} {et~al.}(2008){Verhamme}, {Schaerer}, {Atek}, \&
  {Tapken}}]{verhamme2008}
{Verhamme}, A., {Schaerer}, D., {Atek}, H., \& {Tapken}, C. 2008, \aap, 491,
  89, \dodoi{10.1051/0004-6361:200809648}

\bibitem[{{Verhamme} {et~al.}(2006){Verhamme}, {Schaerer}, \&
  {Maselli}}]{verhamme2006}
{Verhamme}, A., {Schaerer}, D., \& {Maselli}, A. 2006, \aap, 460, 397,
  \dodoi{10.1051/0004-6361:20065554}

\bibitem[{{Verheijen}(2001)}]{verheijen2001}
{Verheijen}, M. A.~W. 2001, \apj, 563, 694, \dodoi{10.1086/323887}

\bibitem[{{Vogt} {et~al.}(1994){Vogt}, {Allen}, {Bigelow}, {Bresee}, {Brown},
  {Cantrall}, {Conrad}, {Couture}, {Delaney}, {Epps}, {Hilyard}, {Hilyard},
  {Horn}, {Jern}, {Kanto}, {Keane}, {Kibrick}, {Lewis}, {Osborne},
  {Pardeilhan}, {Pfister}, {Ricketts}, {Robinson}, {Stover}, {Tucker}, {Ward},
  \& {Wei}}]{Keck-HIRES}
{Vogt}, S.~S., {Allen}, S.~L., {Bigelow}, B.~C., {et~al.} 1994, in Society of
  Photo-Optical Instrumentation Engineers (SPIE) Conference Series, Vol. 2198,
  Instrumentation in Astronomy VIII, ed. D.~L. {Crawford} \& E.~R. {Craine},
  362, \dodoi{10.1117/12.176725}

\bibitem[{{Walter} {et~al.}(2008){Walter}, {Brinks}, {de Blok}, {Bigiel},
  {Kennicutt}, {Thornley}, \& {Leroy}}]{walter2008}
{Walter}, F., {Brinks}, E., {de Blok}, W.~J.~G., {et~al.} 2008, \aj, 136, 2563,
  \dodoi{10.1088/0004-6256/136/6/2563}

\bibitem[{{Walter} {et~al.}(2020){Walter}, {Carilli}, {Neeleman}, {Decarli},
  {Popping}, {Somerville}, {Aravena}, {Bertoldi}, {Boogaard}, {Cox}, {da
  Cunha}, {Magnelli}, {Obreschkow}, {Riechers}, {Rix}, {Smail}, {Weiss},
  {Assef}, {Bauer}, {Bouwens}, {Contini}, {Cortes}, {Daddi}, {Diaz-Santos},
  {Gonz{\'a}lez-L{\'o}pez}, {Hennawi}, {Hodge}, {Inami}, {Ivison}, {Oesch},
  {Sargent}, {van der Werf}, {Wagg}, \& {Yung}}]{walter2020}
{Walter}, F., {Carilli}, C., {Neeleman}, M., {et~al.} 2020, \apj, 902, 111,
  \dodoi{10.3847/1538-4357/abb82e}

\bibitem[{{Wang} {et~al.}(2016){Wang}, {Koribalski}, {Serra}, {van der Hulst},
  {Roychowdhury}, {Kamphuis}, \& {Chengalur}}]{wang2016b}
{Wang}, J., {Koribalski}, B.~S., {Serra}, P., {et~al.} 2016, \mnras, 460, 2143,
  \dodoi{10.1093/mnras/stw1099}

\bibitem[{{Wang} {et~al.}(2015){Wang}, {Kanekar}, \& {Prochaska}}]{wang2015}
{Wang}, W.-H., {Kanekar}, N., \& {Prochaska}, J.~X. 2015, \mnras, 448, 2832,
  \dodoi{10.1093/mnras/stv171}

\bibitem[{{Werle} {et~al.}(2022){Werle}, {Poggianti}, {Moretti}, {Bellhouse},
  {Vulcani}, {Gullieuszik}, {Radovich}, {Fritz}, {Ignesti}, {Richard},
  {Soucail}, {Bruzual}, {Charlot}, {Mingozzi}, {Bacchini}, {Tomicic}, {Smith},
  {Kulier}, {Peluso}, \& {Franchetto}}]{werle2022}
{Werle}, A., {Poggianti}, B., {Moretti}, A., {et~al.} 2022, \apj, 930, 43,
  \dodoi{10.3847/1538-4357/ac5f06}

\bibitem[{{Wetzel} {et~al.}(2013){Wetzel}, {Tinker}, {Conroy}, \& {van den
  Bosch}}]{wetzel2013}
{Wetzel}, A.~R., {Tinker}, J.~L., {Conroy}, C., \& {van den Bosch}, F.~C. 2013,
  \mnras, 432, 336, \dodoi{10.1093/mnras/stt469}

\bibitem[{{Whitaker} {et~al.}(2014){Whitaker}, {Franx}, {Leja}, {van Dokkum},
  {Henry}, {Skelton}, {Fumagalli}, {Momcheva}, {Brammer}, {Labb{\'e}},
  {Nelson}, \& {Rigby}}]{whitaker2014}
{Whitaker}, K.~E., {Franx}, M., {Leja}, J., {et~al.} 2014, \apj, 795, 104,
  \dodoi{10.1088/0004-637X/795/2/104}

\bibitem[{{Wolfe} {et~al.}(2005){Wolfe}, {Gawiser}, \& {Prochaska}}]{wolfe2005}
{Wolfe}, A.~M., {Gawiser}, E., \& {Prochaska}, J.~X. 2005, \araa, 43, 861,
  \dodoi{10.1146/annurev.astro.42.053102.133950}

\bibitem[{{Zwaan} {et~al.}(2005){Zwaan}, {Meyer}, {Staveley-Smith}, \&
  {Webster}}]{zwaan2005}
{Zwaan}, M.~A., {Meyer}, M.~J., {Staveley-Smith}, L., \& {Webster}, R.~L. 2005,
  \mnras, 359, L30, \dodoi{10.1111/j.1745-3933.2005.00029.x}

\end{thebibliography}
\bibliographystyle{aasjournal}



\end{document}